\documentclass[aps,prb,floatfix,twocolumn,showpacs,nofitinbib,hyperref=pdftex,citeautoscript]{revtex4-1}
\usepackage{epsfig}
\usepackage{amsmath,bm}
\usepackage{amssymb}
\usepackage{amsfonts}
\usepackage{dsfont}
\usepackage{comment,color}
\usepackage{lipsum}
\usepackage{hyperref}

\hypersetup{colorlinks=true,linkcolor=blue,anchorcolor=blue,citecolor=blue,filecolor=blue,urlcolor=blue,bookmarksnumbered=true,pdfview=FitB}

\newcommand{\nn}{\nonumber\\}
\usepackage[dvipsnames]{xcolor}

\begin{document}

\title{Transport signatures of topological phases in double nanowires \\ probed by  spin-polarized STM }

\author{Manisha Thakurathi, Denis Chevallier, Daniel Loss, and Jelena Klinovaja}

\affiliation{Department of Physics, University of Basel,
Klingelbergstrasse 82, CH-4056 Basel, Switzerland}

\date{\today}

\begin{abstract}
We study a double-nanowire setup proximity coupled to an $s$-wave superconductor and search for the bulk signatures of the topological phase transition that can be observed experimentally, for example, with an STM tip. Three bulk quantities, namely, the charge, the spin polarization, and  the pairing amplitude of intrawire superconductivity are studied in this work. The spin polarization and the pairing amplitude flip sign as the system undergoes a phase transition from the trivial to the topological phase. In order to identify  promising ways to observe bulk signatures of the phase transition in transport experiments,
we compute the spin current flowing between a local spin-polarized probe, such as  an STM tip, and the double-nanowire system in the Keldysh formalism. We find that the spin current contains  information about the sign flip of the bulk spin polarization and can be used to determine the topological phase transition point. 
\end{abstract}

\maketitle

\section{Introduction}
Majorana bound states (MBSs) have attracted a lot of attention in recent years due to their potential application in topological quantum computing [\onlinecite{Kitaev,Alicea,Beenakker1,review,elsaRW}]. For example, MBSs appear at zero energy and are localized at the ends of the one-dimensional topological superconductor. The promising platforms to engineer topological superconductivity are semiconducting Rashba nanowires (NWs) subjected to a uniform magnetic field [\onlinecite{M1,M2,Mourik,Das,Deng,Liu,Marcus}] or chains of magnetic adatoms [\onlinecite{Yazdani,JK1,Glazman,JK2,Chain1,Chain2,Chain3,Chain4,teemu}].  However, magnetic field and superconductivity have detrimental effects on each other, which has motivated proposals for time-reversal invariant topological superconductors to avoid the need of magnetic fields, particular examples being   double-NW setups with Karmers pairs of MBSs [\onlinecite{JK3,Flensberg,JK4,JK5,CS,Baba,OD,MT22,chris2,Haim,Ferd,Haim2,Hsu,kon,milena}]. In such  setups, two types of proximity induced superconductivity play a crucial role: intrawire ($\Delta$) and interwire ($\Delta_c$) superconductivity. The latter pairing mechanism is also known as crossed Andreev 
reflection
 [\onlinecite{CAR0,CAR1,CAR2,CAR3,CAR4,CAR5,CAR6,CAR7,CAR8,CAR9,bena}].  
 A double-NW setup also reduces the magnetic field required to reach the phase with a single MBS and therefore
 exhibits a richer phase diagram with three phases: trivial phase, phase with one MBS, and a phase with two MBSs.  
 However, to obtain Kramers pairs of MBSs at the end of the system in the absence of a field, strong electron-electron interactions are required such that $\Delta_c>\Delta$ [\onlinecite{chris},\onlinecite{MT1}].  At the same time, a finite value of $\Delta_c$, even if smaller than $\Delta$, is useful since it helps to weaken the requirement on the magnetic field strength needed to enter a phase with one MBS and, moreover, to keep the localization length of the MBS shorter compared to the more common case of a  setup with a single nanowire [\onlinecite{CS}, \onlinecite{OD}]. It is this fact which motivates us to focus on this parameter regime.

\begin{figure}[t]
\includegraphics[width=0.9 \linewidth]{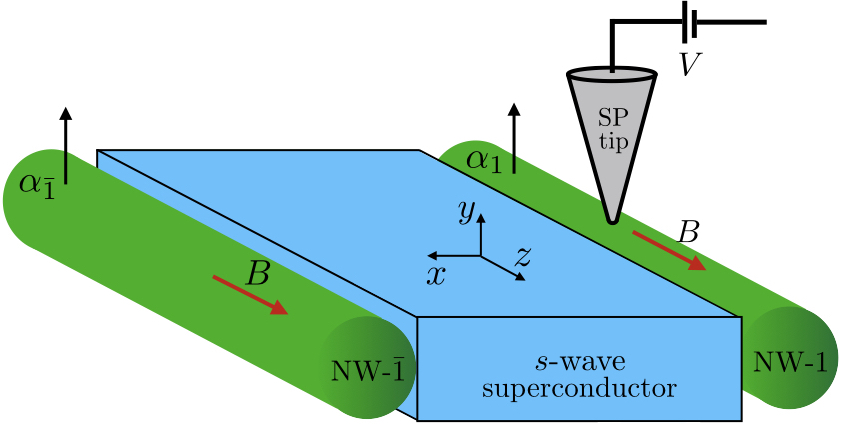}
\caption{Schematics  of the setup consisting of two one-dimensional Rashba NWs (green cylinders) that are aligned along  the $z$ axis and are in proximity to an $s$-wave bulk superconductor (blue slab). An external magnetic field $\bm{B}$ is applied along the axis of NWs and is perpendicular to the Rashba SOI vectors $\bm{\alpha}_1$ and $\bm{\alpha}_{\bar 1}$. The transport simulations are performed for a weakly coupled   spin-polarized STM tip (gray cone) which is biased at the voltage $V$ with respect to the bulk superconductor. The role of the STM can be played by any other local current probe that is spin-selective.}
\label{fig01}
\end{figure} 

Most of the experimental and theoretical work until now is based on the transport signature coming from the MBS  [\onlinecite{Mourik,Das,Deng,Liu,Marcus,Altland,Beenakker2,DL1,MT2, Deb,elsa,pascal,olesia,Fu,Wu,Lu}] rather than a signature coming from the bulk states [\onlinecite{Pawel,Marcel,annica}]. The experimentally observed zero-bias peak is one of the prime signature of MBSs, however, the origin of the peak is ambiguous and can arise from other sources, such as disorder, Kondo resonance, and Andreev bound states [\onlinecite{Brouwer,Ptok,Tewari,Reeg,abs1,abs2,abs3,elsa22}]. Therefore, in this work, we look for the {\it bulk signature} of the topological phase transition and study various bulk properties such as the charge, the spin polarization, and the intrawire pairing amplitude to distinguish between different phases. We numerically calculate these bulk properties and illustrate that the spin projection along the external magnetic field and the intrawire pairing amplitude flip their sign as the system undergoes a the topological phase transition.  There are different ways to measure this sign flip of the spin polarization, for example, by using an STM  or a quantum dot [\onlinecite{Silas,DC_PS,stm1,stm2,stm3,stm4,stm5,DC_CJ,stm6,stm7,DC,elsa2,ali1}]. Each approach has its own advantages and disadvantages. For instance, creating a quantum dot at the end of the NW allows one to perform the spectroscopy of the NW and filter both the spin and  energy of the transported electrons by properly choosing the size of the quantum dot. However, it is not possible to study the spatial dependence of the current in this case. 
In contrast, an STM tip is highly controllable and allows local measurements.
 For example, if one is interested in the MBSs (bulk states), one positions the tip at the end (middle) of the NW. In this work, with aforementioned advantages, we demonstrate that there is a detectable  sign flip of the spin polarization of the lowest band when using 
a spin-polarized local current probe such as a spin-polarized (SP) STM tip [\onlinecite{ali,kat,ham}].
The computed spin current flips sign exactly when the component of the spin polarization along the external magnetic field flips sign
  as we show in  numerical simulations based on the Keldysh Green function formalism [\onlinecite{D1},\onlinecite{D2}].
 
The outline of the paper is as follows.  In Sec. \ref{model}, we discuss the setup under consideration. In Sec. \ref{bulk_quant}, we compute the spectrum and bulk quantities, and in Sec. \ref{SP} we compute numerically the spin current through a weakly coupled spin-polarized STM tip. We conclude in Sec. \ref{con}. Technical details are deferred to two  appendices.

\section{Model}
\label{model}
We consider a double-NW setup shown in Fig. \ref{fig01}, where the NWs are oriented along the $z$ direction and are in proximity with an $s$-wave superconductor. The Rashba spin-orbit interaction (SOI) vector is pointing along the $y$ direction in both NWs. The kinetic part of the Hamiltonian has the following form:
\begin{align}
&H_{kin} =\sum_\eta \int dz\ \Big[ \sum_{\sigma} c_{\eta\sigma}^\dagger(z) \left( \frac{-\hbar^2 \partial_z^2}{2m_0}  - \mu_\eta \right)c_{\eta\sigma}(z) \nn
&\hspace{55 pt}- i \sum_{\sigma,\sigma'} \alpha_{\eta} \, c_{\eta\sigma}^\dagger(z) \,(\sigma_y)_{\sigma\sigma'} \,\partial_z\,c_{\eta\sigma'}(z)\Big],
\end{align}
where $c^\dagger_{\eta\sigma}(z)$
creates an electron with effective mass $m_0$ and spin $\sigma$ at position $z$ in the NW $\eta$. For the $\eta$-NW, the strength of the SOI is given by $\alpha_\eta$  which is related to the SOI momentum by $k_{so,\eta}=m_0\alpha_\eta/\hbar^2$.  The chemical potential is given by $\mu_\eta$. 	Without loss of generality,  we  consider $\alpha_1>\alpha_{\bar 1}$ [\onlinecite{JK3}]. The proximity induced superconductivity is described by the Hamiltonian
\begin{align}
H_{sc}= &\sum_{\eta,\sigma,\sigma'} \int dz\,  \Big[ \frac{\Delta_\eta}{2}
 c_{\eta\sigma}(z) \, (i \, \sigma_y)_{\sigma\sigma'}\, c_{\eta\sigma'}(z) \nn
&\hspace*{1.2cm}+\frac{\Delta_{c}}{2} 
 c_{\eta\sigma}(z)\, (i \, \sigma_y)_{\sigma\sigma'} \, c_{\bar \eta\sigma'}(z)+  \text{H.c.}\,\Big],
\label{HSC}
\end{align}
where the first (second) term is the intrawire (interwire) superconductivity with pairing amplitudes $\Delta_\eta$ ($\Delta_c$) corresponding to the process when the two electrons in the Cooper pair tunnel into the same NW (different NWs). We set
the interwire single-electron  tunneling to zero as it has been shown in previous work that its effect can be compensated by tuning the chemical potential to a sweet spot [\onlinecite{CS}]. Therefore, all results obtained in the following sections, are valid also for the case of finite interwire tunneling. Moreover, the setup is subjected to an external magnetic field $B$ along the NW, leading to a Zeeman energy $\Delta_{ Z\eta}=g_\eta\mu_B B/2$ where $g_\eta$ corresponds to the $g$-factor of the $\eta$-NW. Orbital magnetic effects are neglected [\onlinecite{or1,or2,or3,or4,or5}]. The corresponding Hamiltonian reads
\begin{align}
H_Z=\sum_{\eta,\sigma,\sigma'} \Delta_{Z\eta} \int dz \,c_{\eta \sigma}^\dagger(z) \,(\sigma_z)_{\sigma \sigma'} \,c_{\eta \sigma}(z).
\end{align}

\begin{figure}[t]
\begin{center}
\includegraphics[width=1 \linewidth]{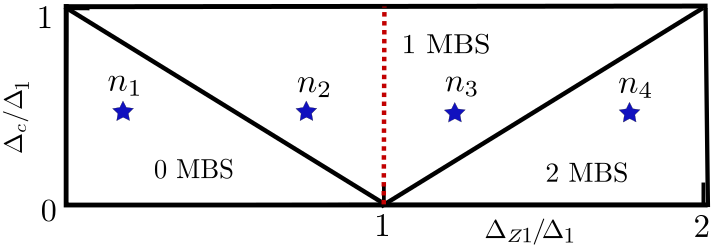}
\end{center}
\caption{Topological phase diagram as a function of the Zeeman splitting $\Delta_{Z1}$ and the interwire pairing amplitude $\Delta_c$. The black solid line divides the phase diagram into three phases, namely trivial phase and topological phases with one MBS and two MBSs. The red line corresponds to a crossover point between the first and second lowest energy band of the double-NW system.  For simplicity, we assume that $\Delta_{Z\bar1}= \Delta_{Z1}$ and $\Delta_{\bar1}= \Delta_{1}$. Different points in the phase diagram are denoted by stars and labeled by $n_1,n_2,n_3,$ and $n_4$, which will be referred to later.}
\label{fig02}
\end{figure}

\begin{figure}[t]
\begin{center}\begin{tabular}{c}
\includegraphics[width=0.9 \linewidth]{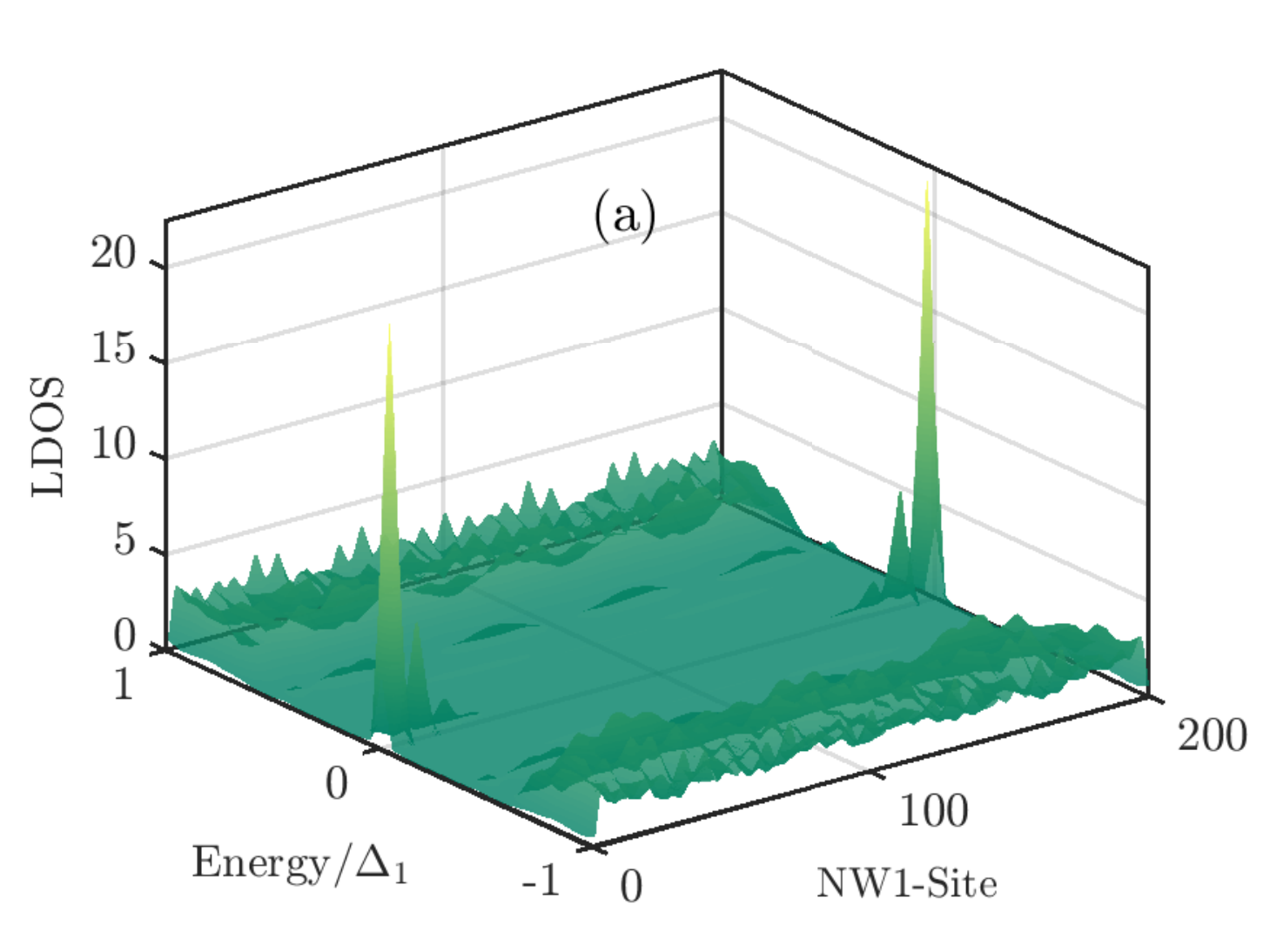}\\
\includegraphics[width=0.9 \linewidth]{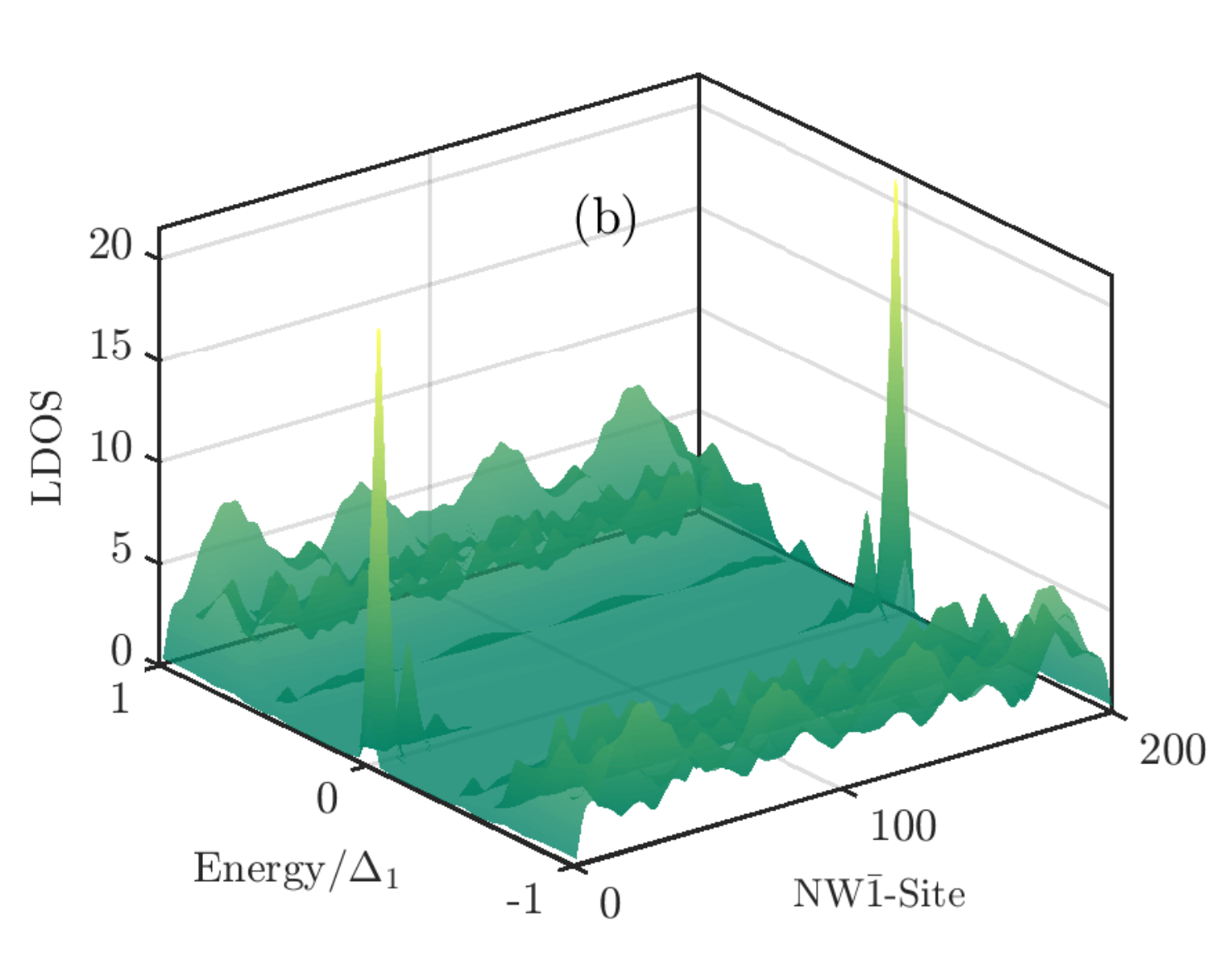}
\end{tabular}
\end{center}
\caption{To confirm the presence of MBSs, we plot the LDOS as a function of energy and position along (a) NW-1 and (b) NW-$\bar 1$ at point $n_2$ of the phase diagram shown in Fig. \ref{fig02}. The single MBS is located in both NWs and clearly visible in the LDOS plot at zero energy. However, the bulk LDOS is slightly different in the panels (a) and (b) due to the unequal strength of the Rashba SOI. Other parameters are $N=200$, $\alpha_1/\alpha_{\bar 1}=1.4$,  
 $E_{so,1}/\Delta_1=1.225$, $\Delta_{\bar 1}/\Delta_1=1$, $\Delta_c/\Delta_{ 1}=0.5$, $ \mu=0$, $\gamma/\Delta_1=0.01$, $\Delta_Z/\Delta_1=0.75$ at point $n_2$.}
\label{fig03}
\end{figure}

 \begin{figure*}[htb]
\begin{center} \begin{tabular}{cccc}
\epsfig{figure=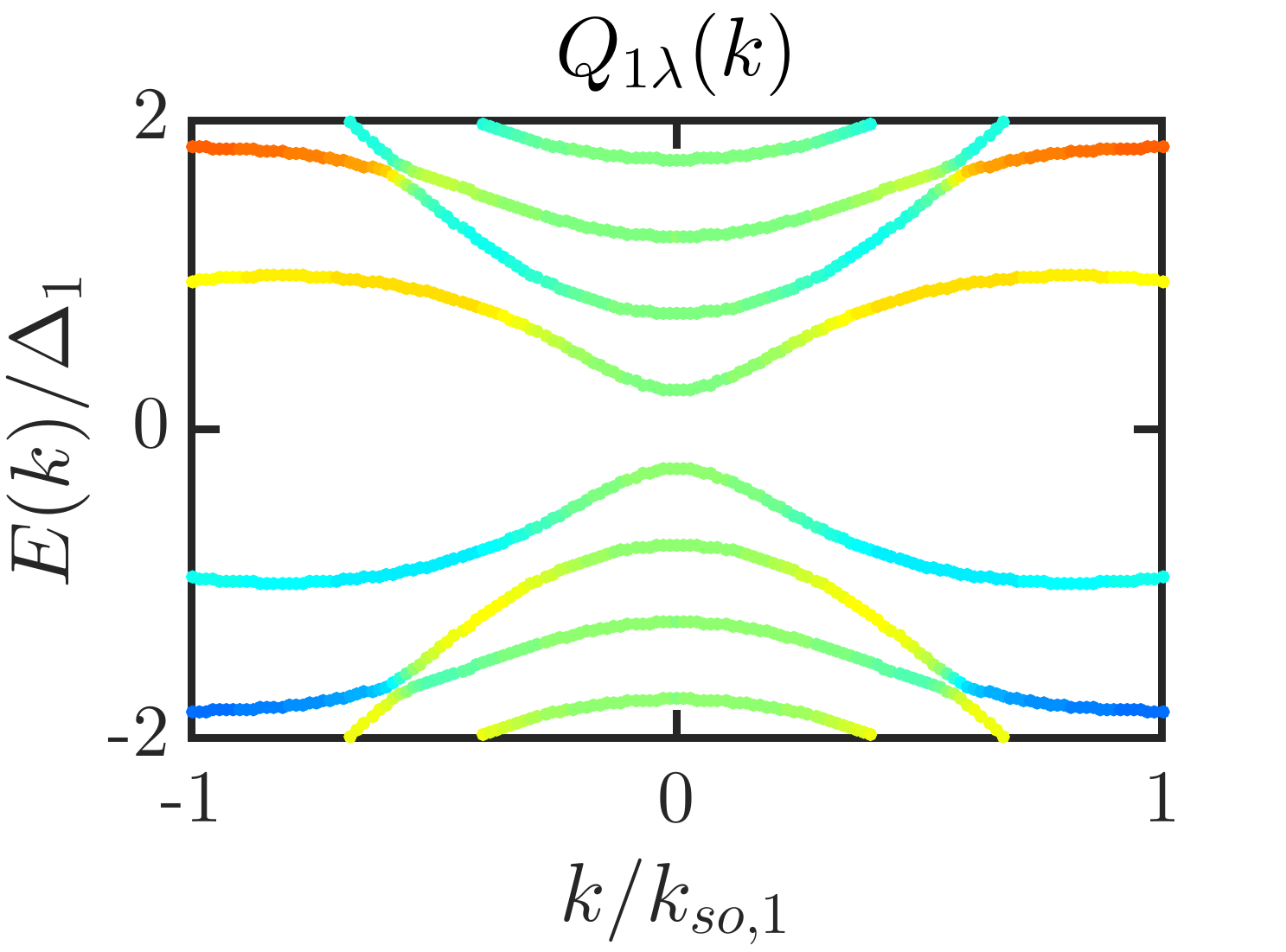,width=1.85in,height=1.45in,clip=true} &\hspace*{-0.5cm}
\epsfig{figure=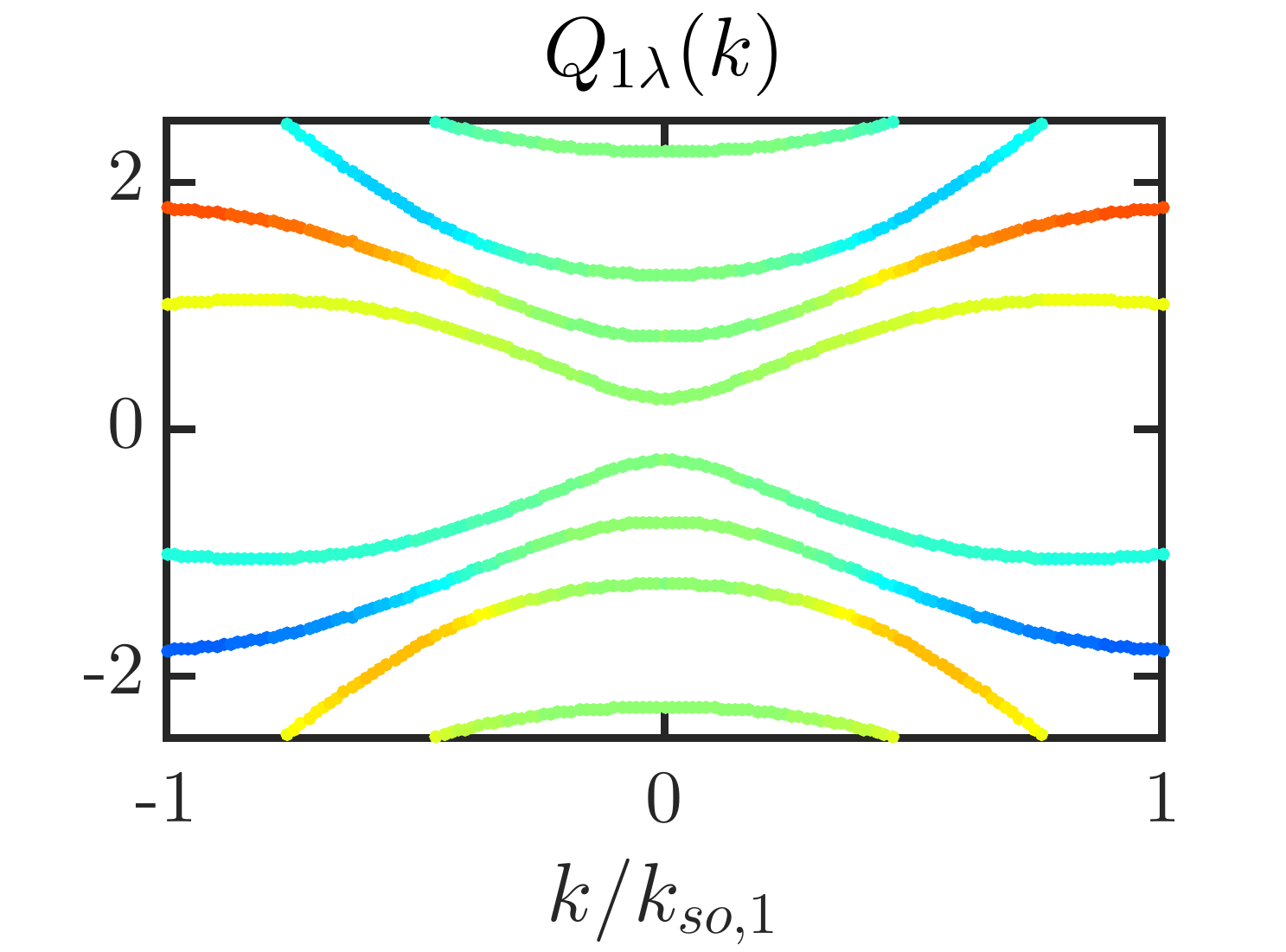,width=1.75in,height=1.45in,clip=true} &\hspace*{-0.5cm}
\epsfig{figure=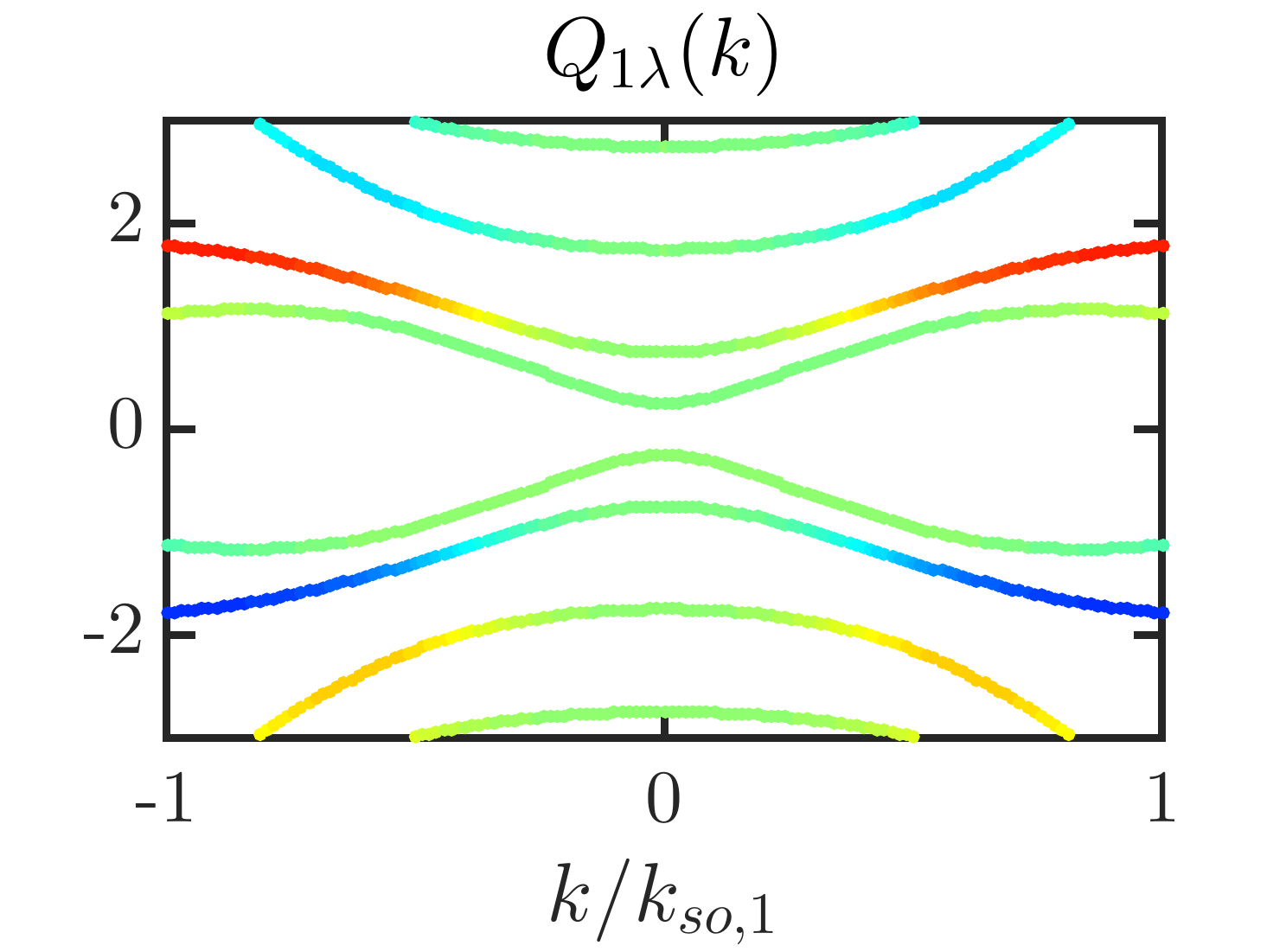,width=1.75in,height=1.45in,clip=true} &\hspace*{-0.5cm}
\epsfig{figure=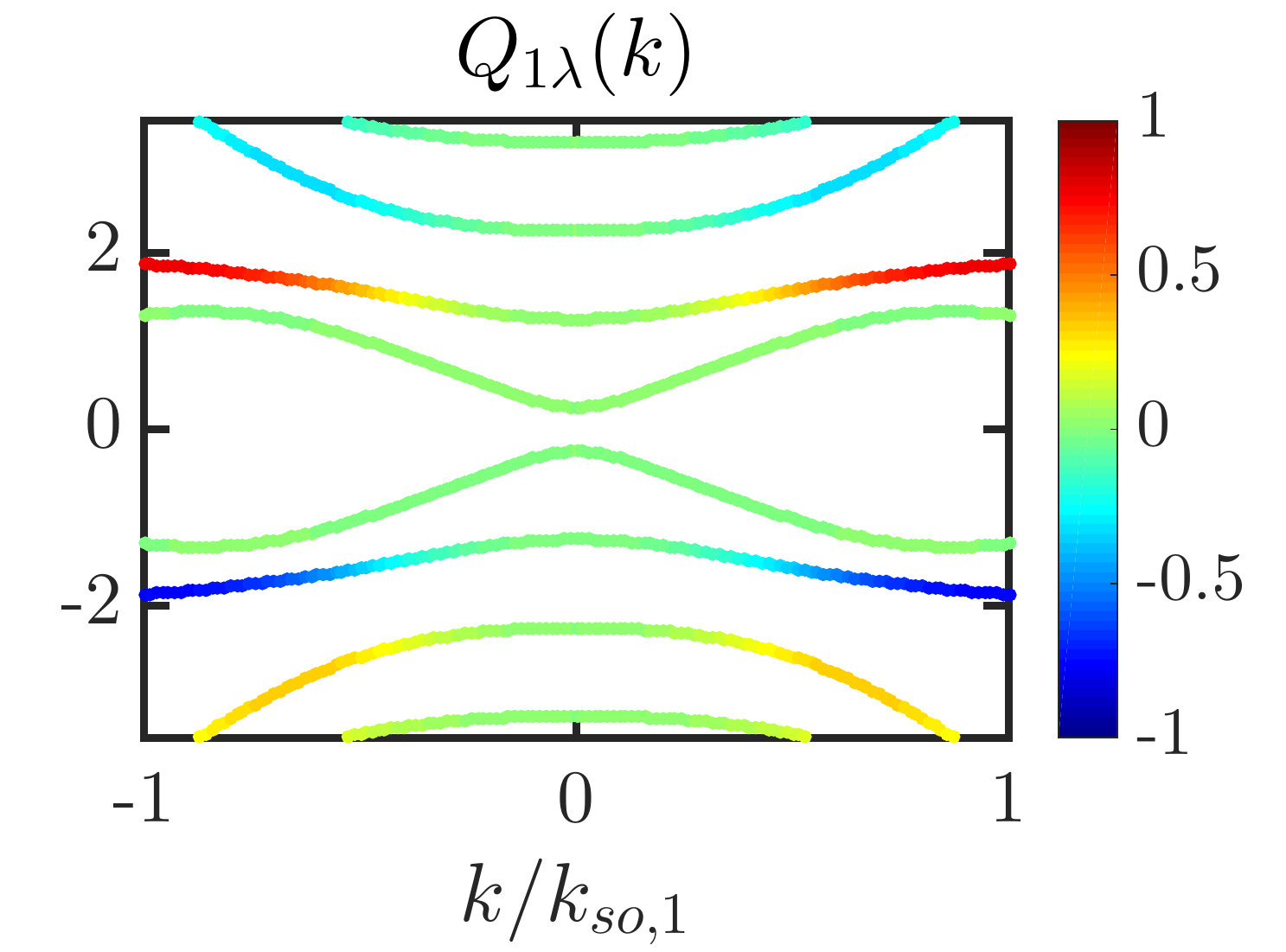,width=1.9in,height=1.45in,clip=true} \\
\epsfig{figure=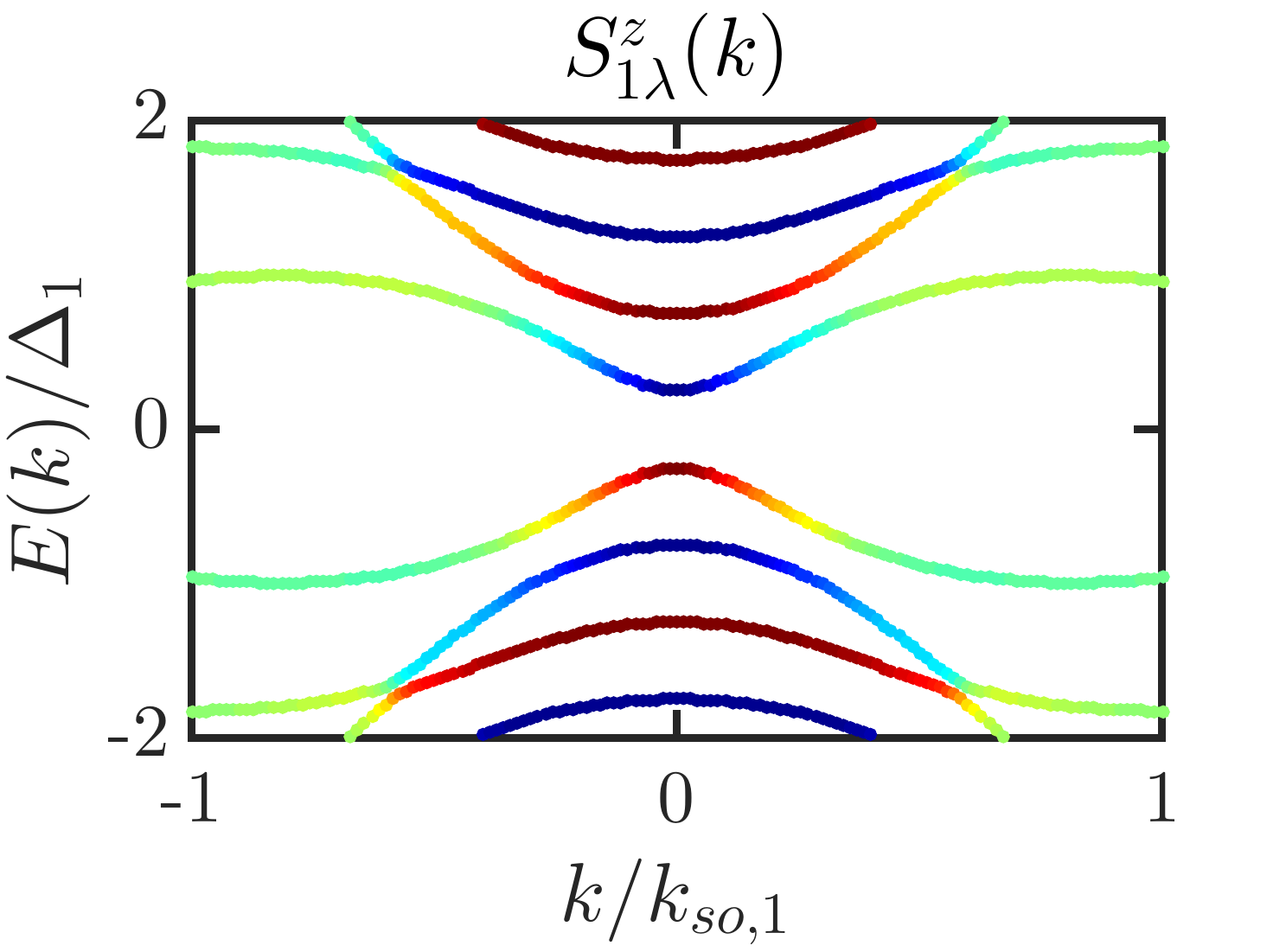,width=1.85in,height=1.45in,clip=true} &\hspace*{-0.5cm}
\epsfig{figure=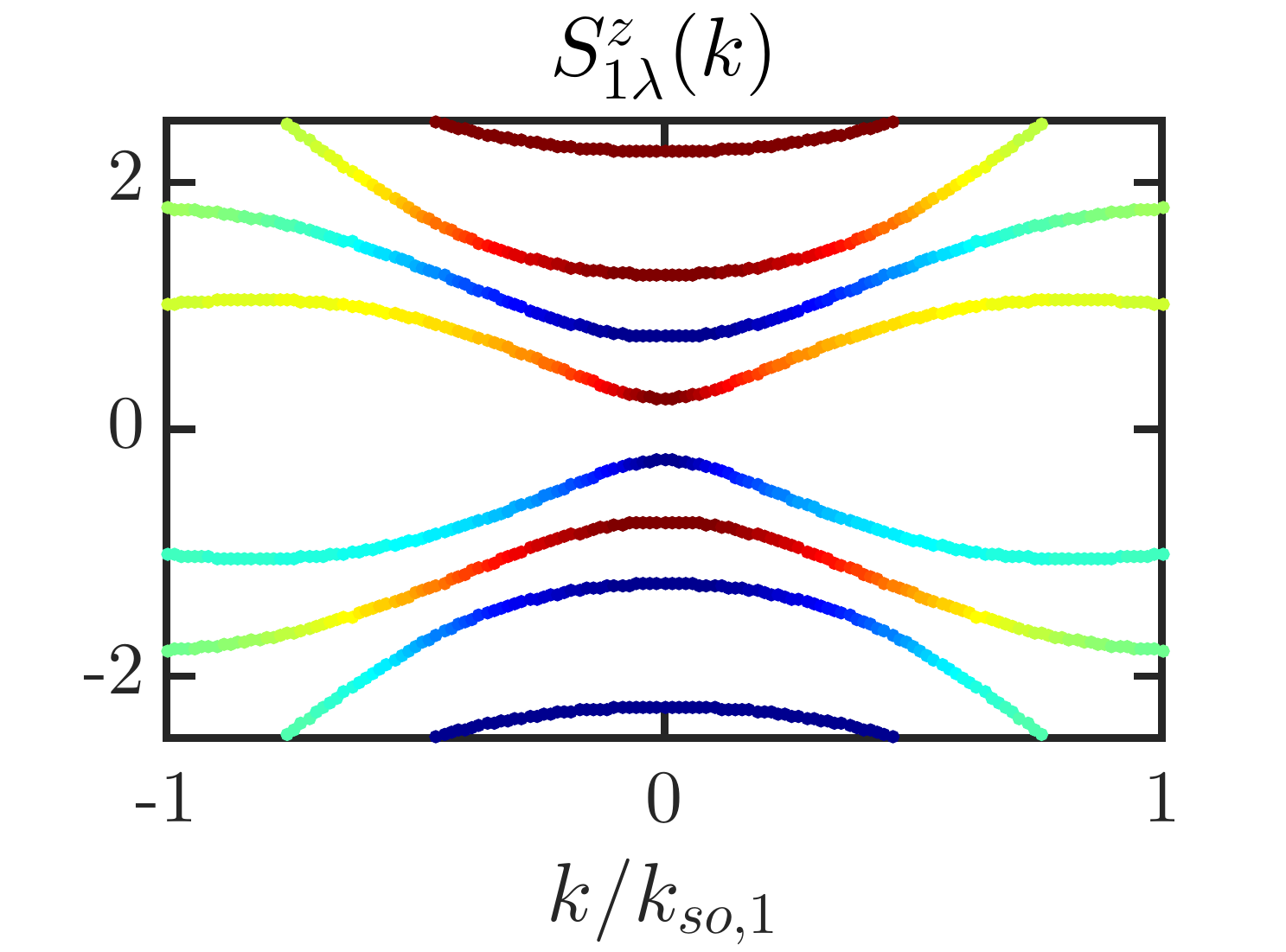,width=1.75in,height=1.45in,clip=true} &\hspace*{-0.5cm}
\epsfig{figure=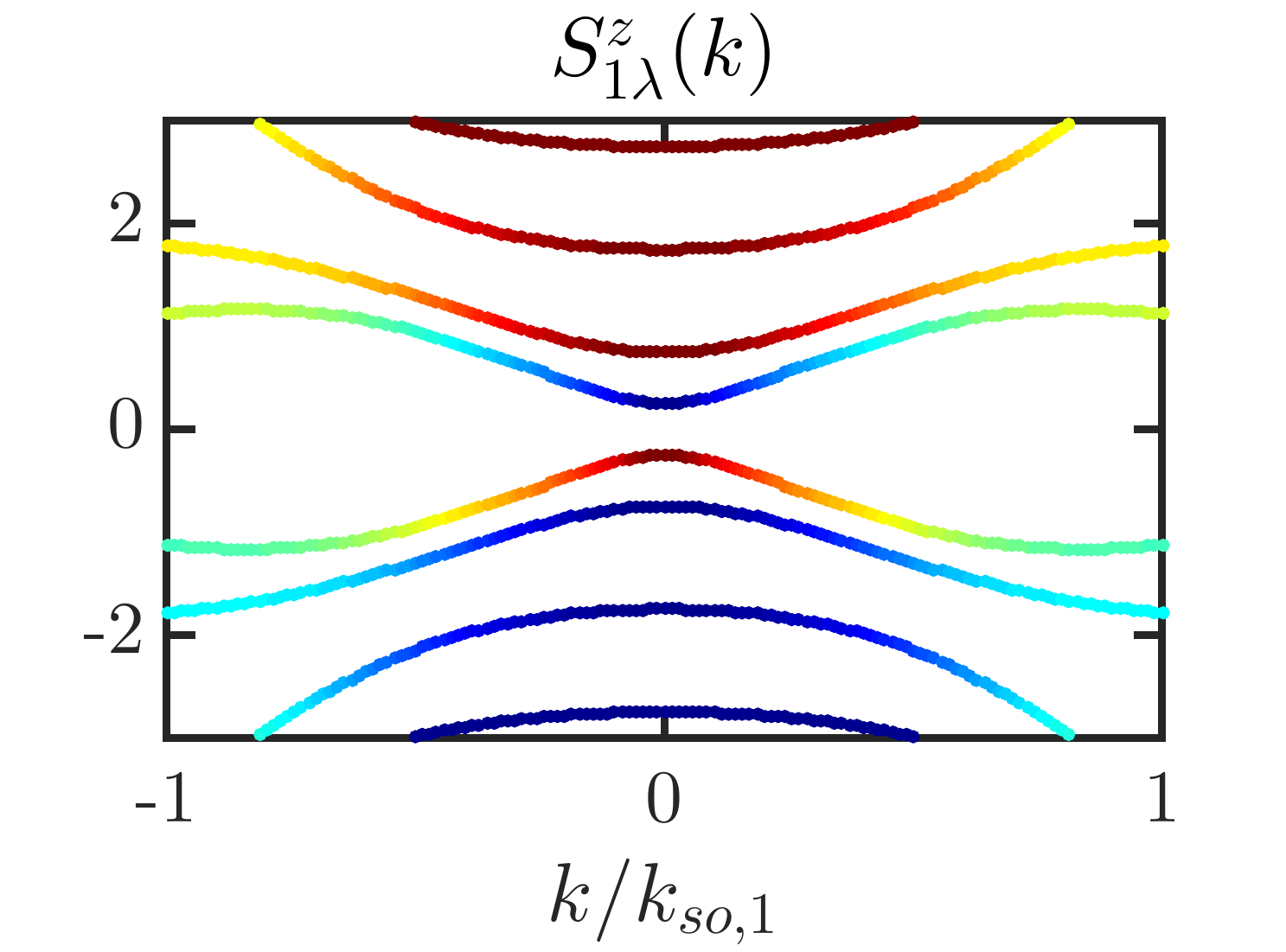,width=1.75in,height=1.45in,clip=true} &\hspace*{-0.5cm}
\epsfig{figure=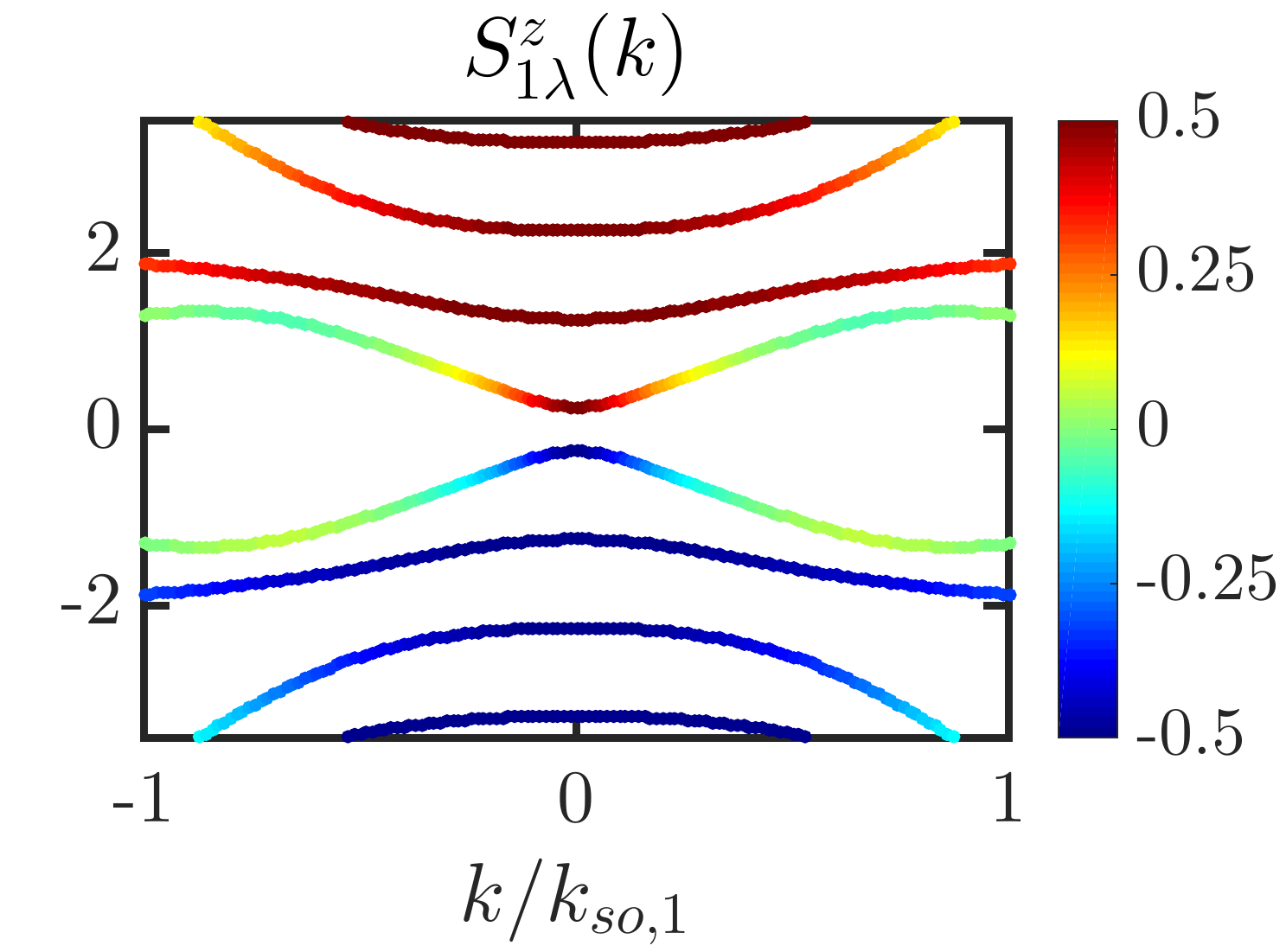,width=1.9in,height=1.45in,clip=true} \\
\epsfig{figure=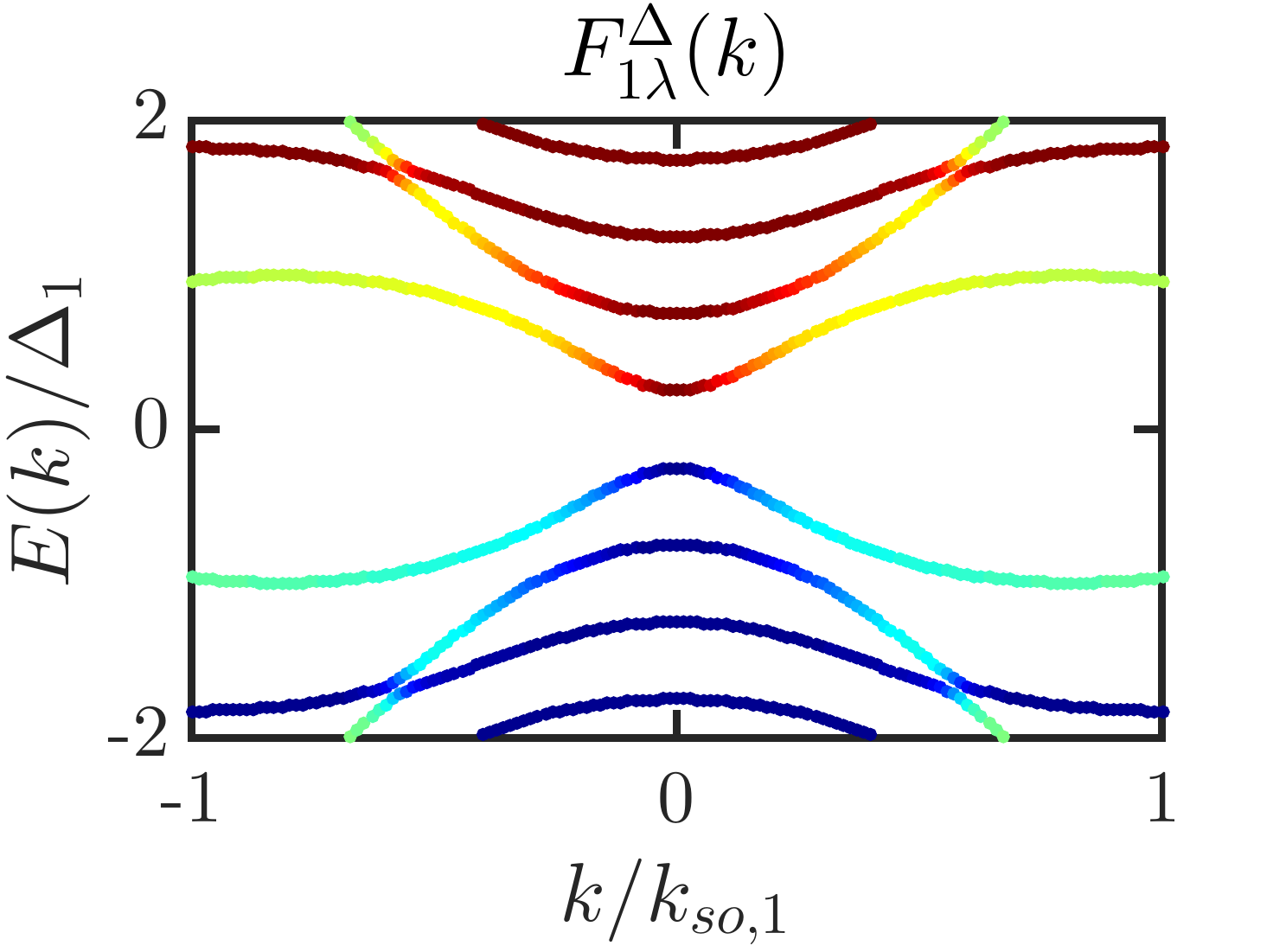,width=1.85in,height=1.45in,clip=true} &\hspace*{-0.5cm}
\epsfig{figure=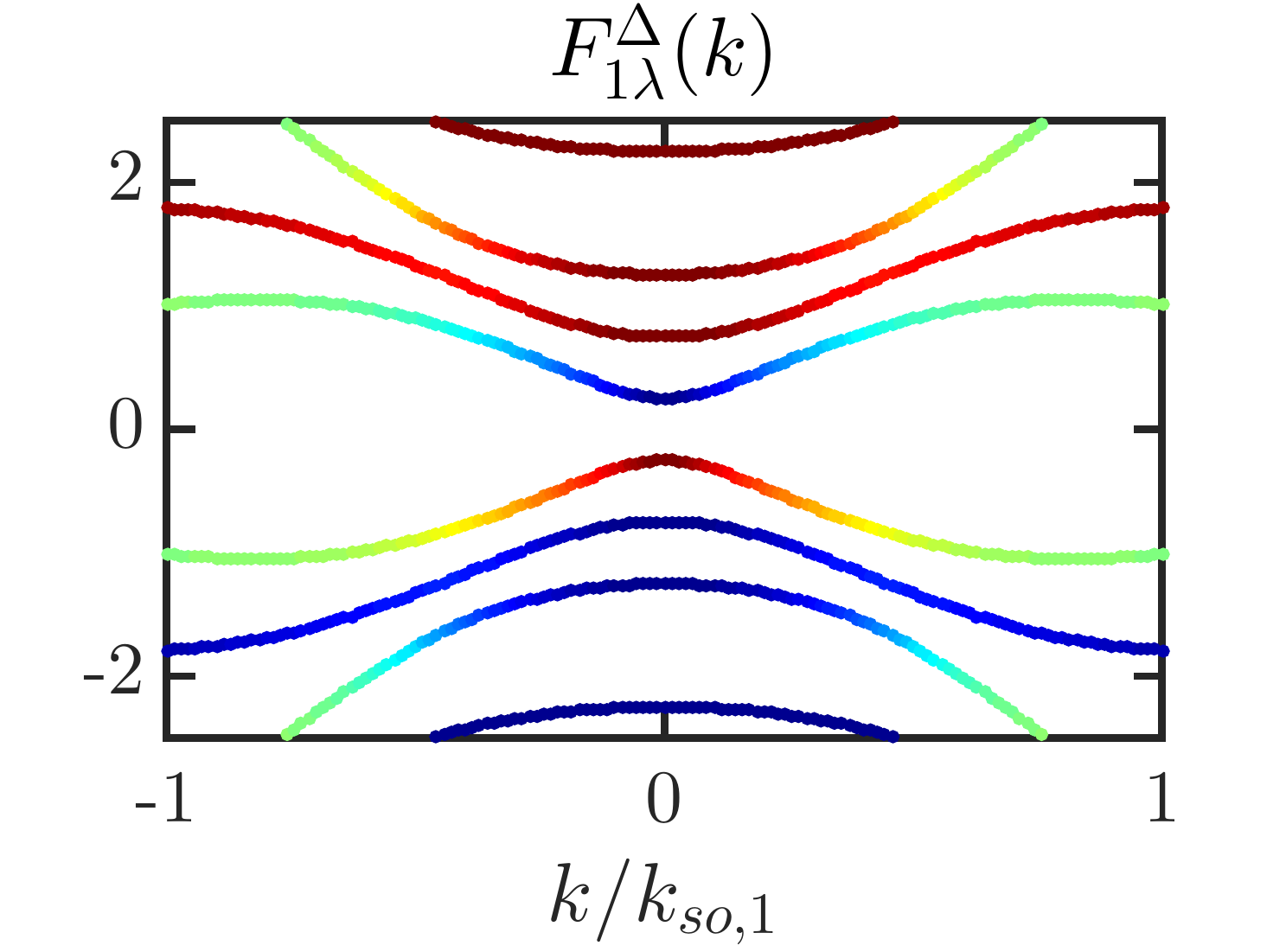,width=1.75in,height=1.45in,clip=true} &\hspace*{-0.5cm}
\epsfig{figure=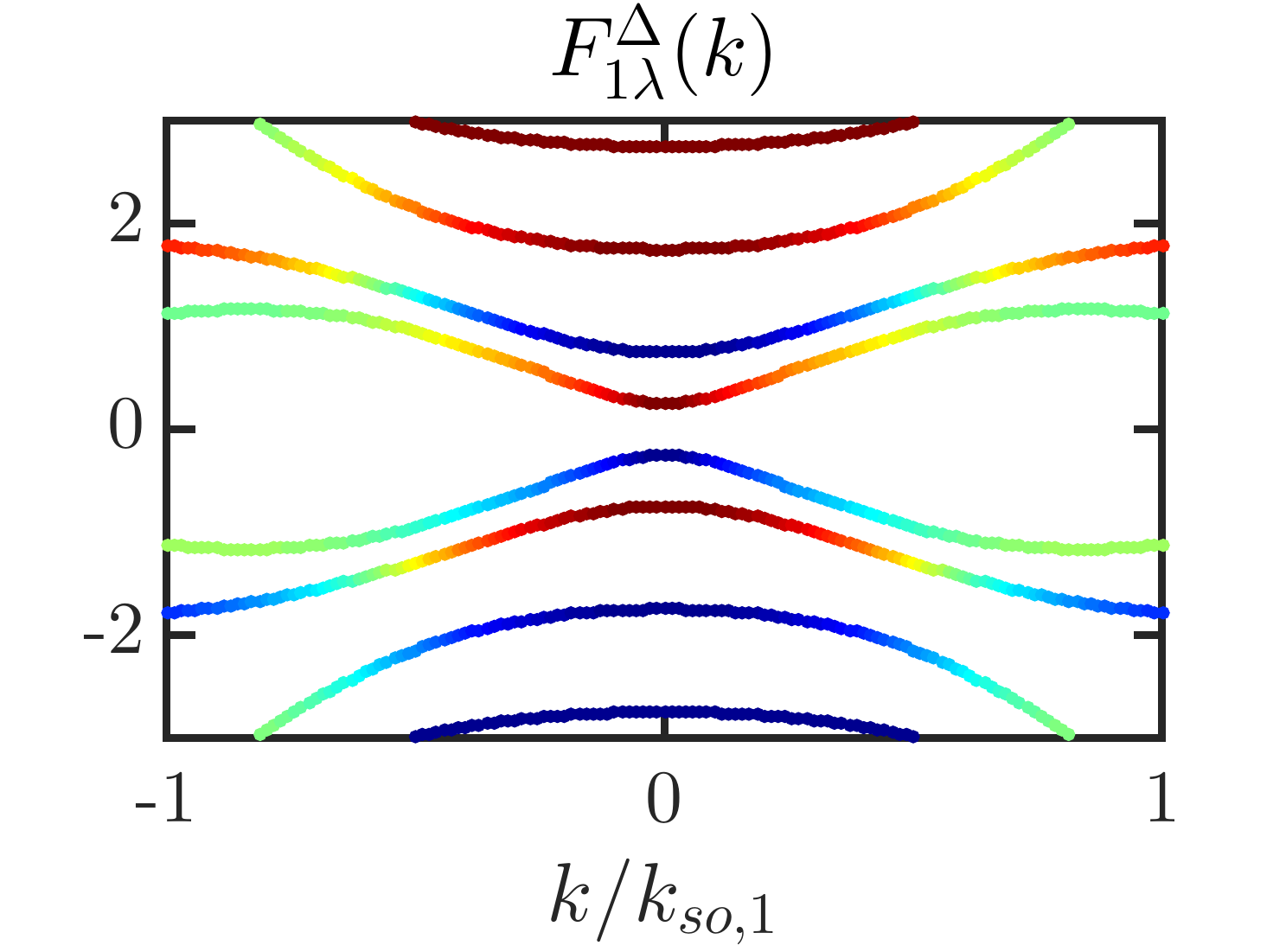,width=1.75in,height=1.45in,clip=true} &\hspace*{-0.5cm}
\epsfig{figure=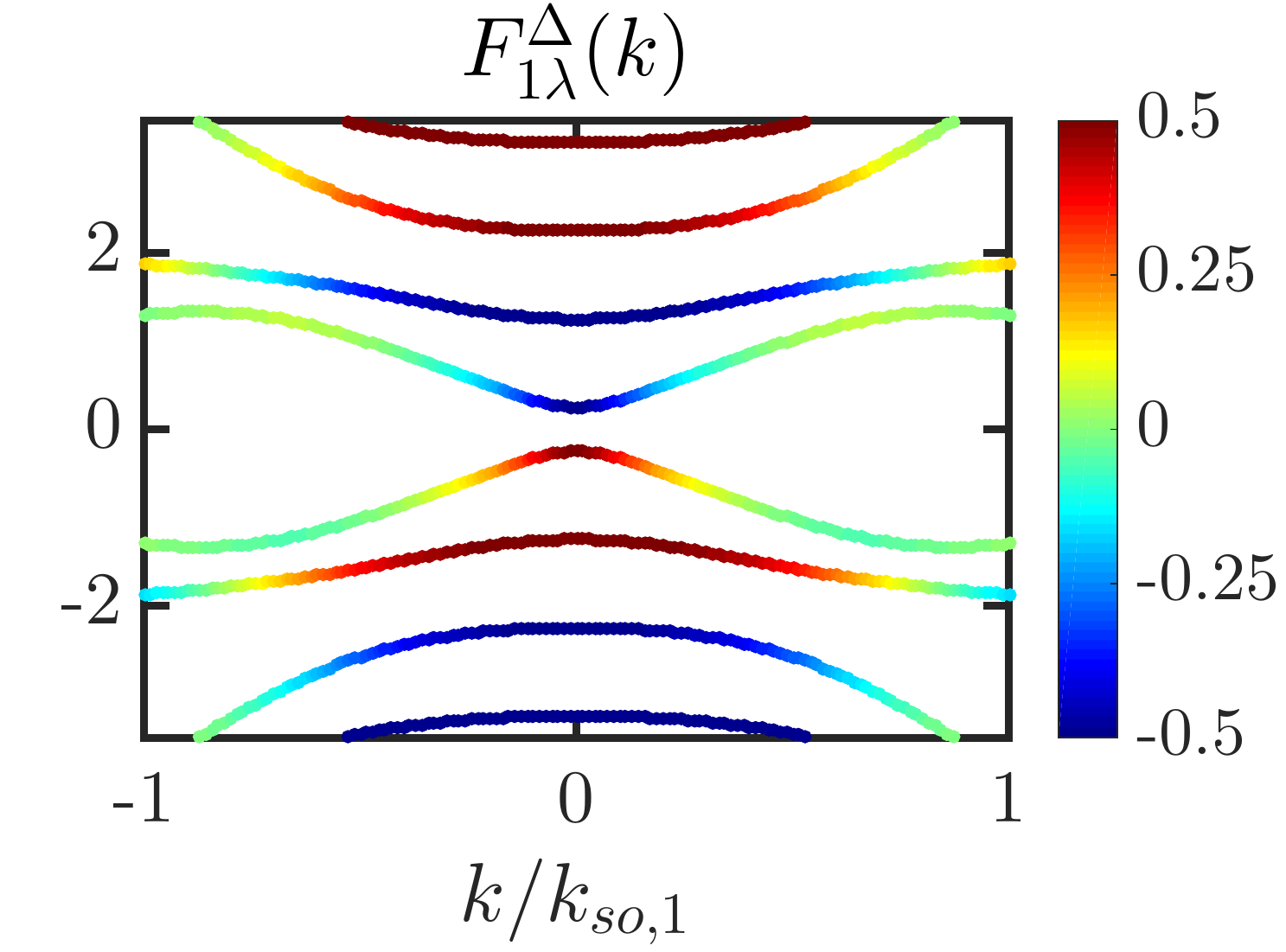,width=1.9in,height=1.45in,clip=true}
\end{tabular} \end{center}
\caption{Energy spectrum for the double-NW system as a function of the momentum $k$ in the setup with periodic boundary conditions [see  Eq. (\ref{H})]. The red and blue colors corresponds to positive and negative values of the bulk quantities - of the charge $Q_{\eta \lambda}$, of the $z$ component of spin polarization $S^z_{\eta \lambda}$, and of the intrawire pairing amplitude $F^\Delta_{\eta\lambda} $ in the row one, two and three, respectively. We plot  $Q_{\eta \lambda}$, $S^z_{\eta \lambda}$, and  $F^\Delta_{\eta\lambda} $ for NW-1 at four different points $n_1-n_4$ in the phase diagram as we go from left to right in each row. The sign flip of the $S^z_{\eta \lambda}$ and $F^\Delta_{\eta\lambda} $ can be clearly seen close to momentum $k=0$. We note that similar behaviour is also obtained for NW-$\bar 1$. Other parameters are chosen as $\Delta_c/\Delta_1=0.5$, $\mu=0$, $\alpha_1/\alpha_{\bar 1}=1.4$, $E_{so,1}/\Delta_1=1.225$, and the points $n_1$, $n_2$, $n_3$, and $n_4$ correspond to $\Delta_Z/\Delta_1= 0.25$, $0.75$, $1.25$, and $1.75$, respectively. }
\label{fig04}
\end{figure*}

To begin with, we consider the continuum limit and work in the basis $\Phi(z)=[c^\dagger_{1\uparrow},c^\dagger_{1\downarrow},c_{1\downarrow},-c_{1\uparrow},c^\dagger_{\bar1 \uparrow},c^\dagger_{\bar 1\downarrow},c_{\bar 1\downarrow},-c_{\bar1\uparrow}]$, in which the total Hamiltonian takes the following form,
\begin{align}
H_0=H_{kin}+H_{sc}+H_Z=\frac{1}{2}\int dz \,\Phi^\dagger(z)\,\mathcal{H}_0(z)\,\Phi(z).
\label{H}
\end{align}
Here, the Hamiltonian density $\mathcal{H}_0(z)$ is given by  
\begin{align}
\mathcal{H}_0= &\left( \frac{\hbar^2\,\hat k^2}{2\,m_0}-\mu_\eta\right)\tau_z+ \alpha_1 \hat k (1+\eta_z) \,\tau_z\,\sigma_y/2\nn
&\hspace{10pt}+\alpha_{\bar 1} \hat k (1-\eta_z)\tau_z\,\sigma_y/2
+\Delta_1 (1+\eta_z)\tau_x/2 \nn
&\hspace{10pt}+ \Delta_{\bar 1} (1-\eta_z) \tau_x/2+\Delta_c\,\eta_x\, \tau_x+\Delta_{Z\eta}\sigma_z,
\label{H0}
 \end{align}
where $\hat k=-i\, \partial_z$  is the momentum operator with the eigenvalue $k$ for bulk eigenstates. The Pauli matrices $\eta_i$, $\tau_i$, and $\sigma_i$ act in the wire, particle-hole, and spin spaces, respectively. 

First, we calculate the phase diagram as a function of the magnetic field $B$ and the interwire pairing amplitude $\Delta_c$ [see Fig. (\ref{fig02})].  The  bulk gap closes at $k=0$ when $\Delta_c^2= (\Delta_{Z1}\pm \Delta_1)^2$. Here, to simplify the expressions, we assume identical NWs with $\Delta_1=\Delta_{\bar 1}$ and $\Delta_{Z1}=\Delta_{Z\bar 1}$ [\onlinecite{CS}]. The energy spectrum of the lowest band near $k=0$ for $\Delta_{Z1}>0$ can be  easily computed from Eq. (\ref{H0}) and is either given  by $E_1= |\Delta_{Z1}-\Delta_1+\Delta_c|$ or by $E_2=|\Delta_{1}-\Delta_{Z1}+\Delta_c|$.  The phase diagram consists of the trivial phase without MBSs and topological phases with one MBS or two MBSs at each end of the setup. It is also important to note that $E_1=0$ ($E_2=0$) corresponds to the topological phase transition point indicating change from zero to one MBS (from one MBS to two MBSs). In addition, these is also a crossover between these two bands, $E_1=E_2$, at $\Delta_{Z1}= \Delta_1$ (shown with a red dashed line in Fig. \ref{fig02}).  This flip between two bands can also be seen in transport experiments as we show below. We also note here that the two MBS phase is present only due to the additional symmetry in the effective model and, thus, it is not stable against arbitrary type of disorder that can be present in the setup  [\onlinecite{CS}]. However, for simplicity, we still refer to it as to the two MBS topological phase to distinguish this region of the topological phase diagram from the zero MBS region in Fig. (\ref{fig02}).

Next, we write the lattice model of the double-NW setup given by the following tight-binding Hamiltonian
\begin{align}
H_{0t}=&\sum_\eta\bigg(\sum_{j=1}^{N} \Phi_{\eta j}^\dagger[-(\mu_\eta-2t) \tau_z+\Delta_\eta \tau_x+\Delta_{Z\eta}\sigma_z] \Phi_{\eta j} \nn
&+\sum_{j=1}^{N-1} \Phi_{\eta\, j+1}^\dagger(-t-i\bar\alpha_\eta \sigma_y)\tau_z \Phi_{\eta j}+\text{H.c.}\bigg)\nn
&+ \sum_{j=1}^N \Phi_{\bar 1 j} (\Delta_c \tau_x) \Phi_{1 j}+\text{H.c.}.
\label{H0t}
\end{align}
where $\Phi_{\eta j}=(c^\dagger_{\eta j1},c^\dagger_{\eta j\bar 1},
c_{\eta j\bar 1},-c_{\eta j1})$ is the electron spinor consisting of the creation operators $c^\dagger_{\eta j\bar \sigma}$ acting on an electron with spin  $\sigma$ at site $j$ of the $\eta$-NW. The spin-conserving hopping amplitude is given by $t=\hbar^2/2m_0a^2$, where $a$ is the lattice spacing. The spin-flip hopping amplitude $\bar \alpha_\eta$  is related to the SOI strength,  $\bar \alpha_\eta=\alpha/2a$, where  $E_{so,\eta}=m_0\alpha_{\eta}^2/2\hbar^2=\bar \alpha^2_\eta/t$ is the SOI energy. In our numerical simulations, we set the hopping amplitude $t=1$, which sets the energy scale for the calculation.  We confirm the presence of zero-energy modes (MBSs) by calculating the local density of state (LDOS) as a function of position and energy [see Fig. \ref{fig03}] given by the following expression:
\begin{align}
\rho_j(\omega)=-\frac{1}{\pi}\sum_\sigma \text{Im}[G_{0R}(\omega)]_{jj,\sigma\sigma},
\end{align}
where 
$\omega$ is the frequency
and  $G_{0R/A}(\omega)=(\omega\pm i\gamma-H_{0t})^{-1}$  the retarded/advanced Green function for the setup, with an infinitesimally small real $\gamma$ required to invert the matrix.

\begin{figure*}[!bt]
\begin{center} \begin{tabular}{cccc}
\epsfig{figure=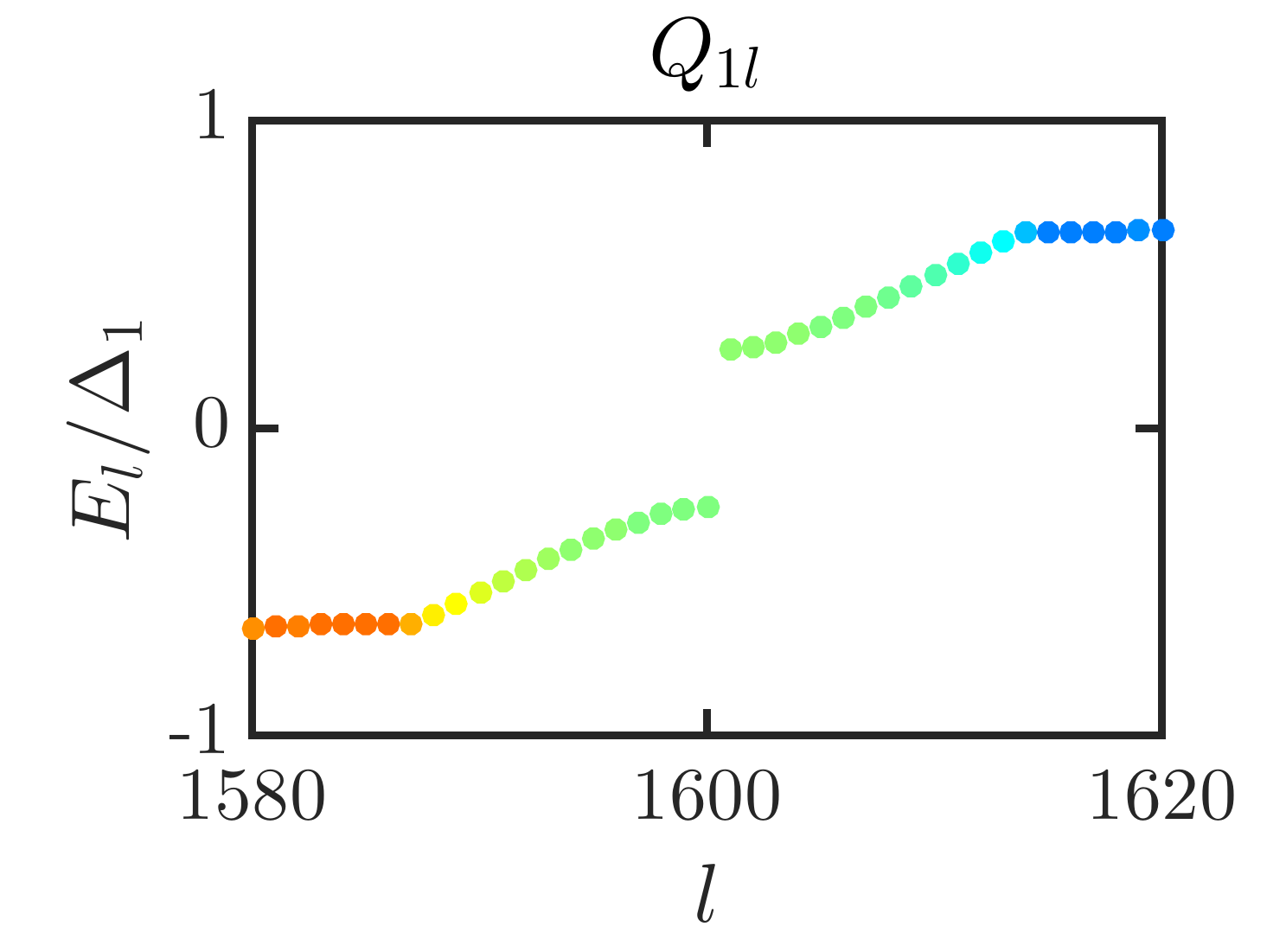,width=1.85in,height=1.45in,clip=true} &\hspace*{-0.4cm}
\epsfig{figure=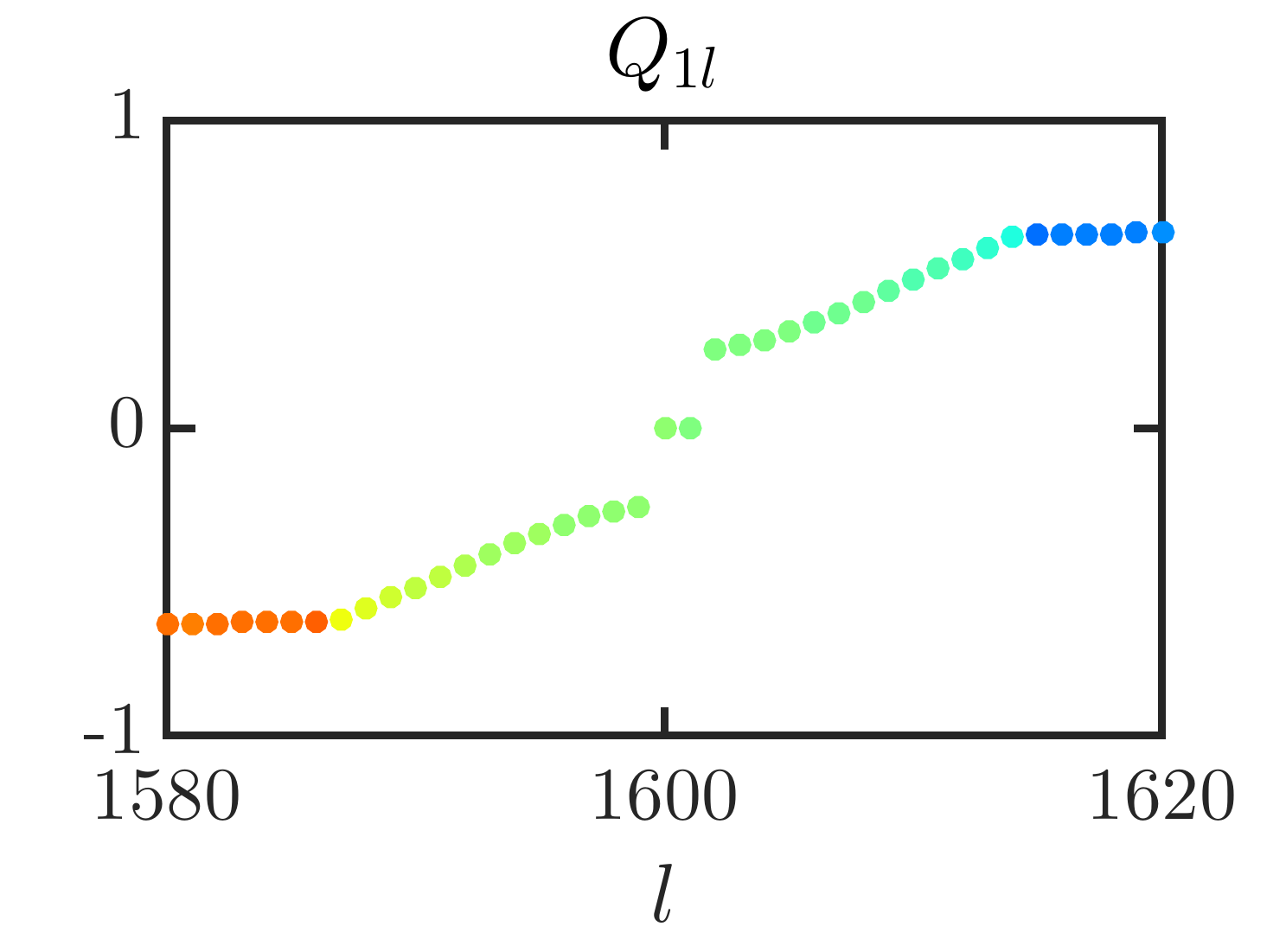,width=1.75in,height=1.45in,clip=true} &\hspace*{-0.4cm}
\epsfig{figure=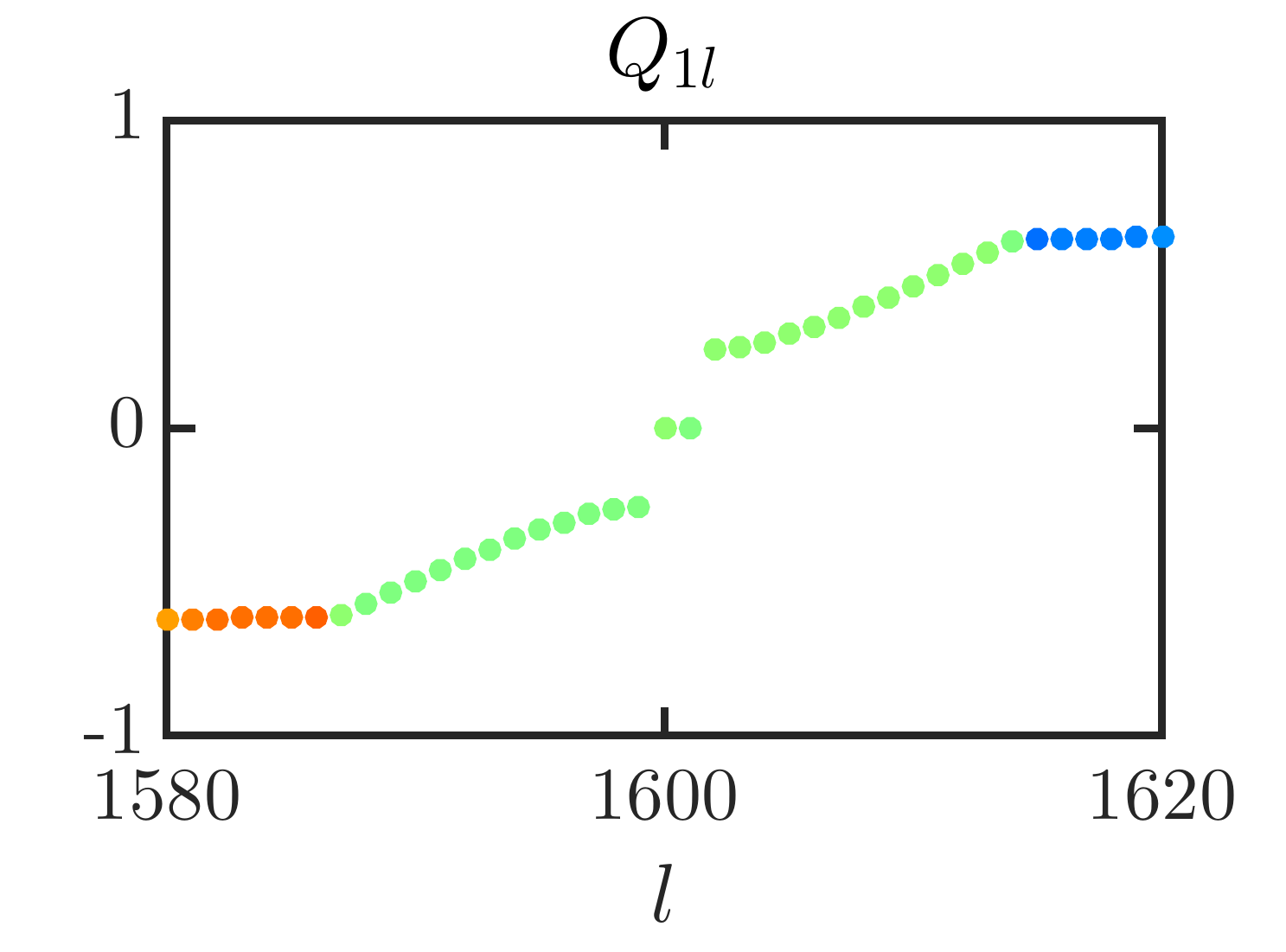,width=1.75in,height=1.45in,clip=true} &\hspace*{-0.4cm}
\epsfig{figure=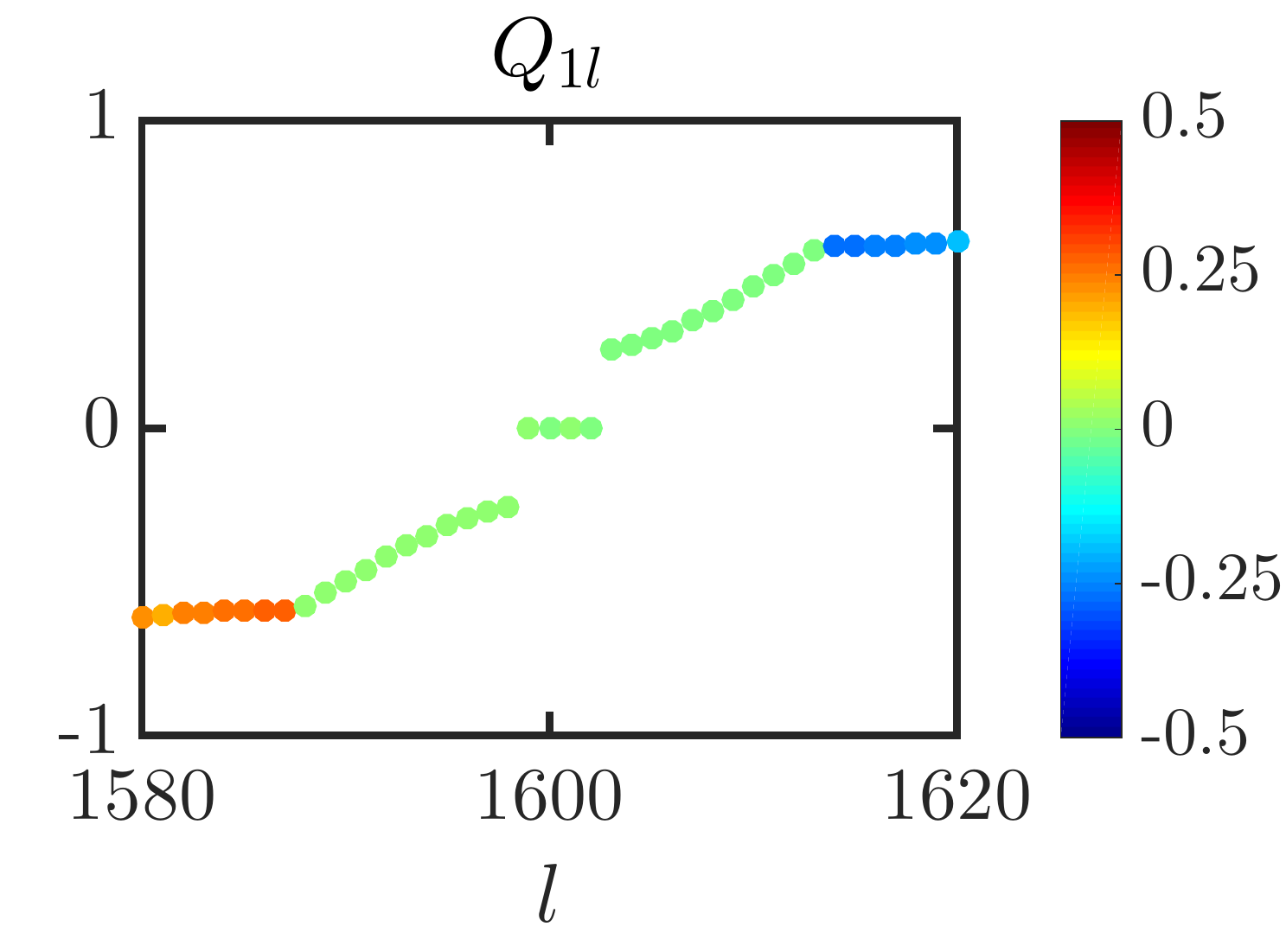,width=1.85in,height=1.45in,clip=true} \\
\epsfig{figure=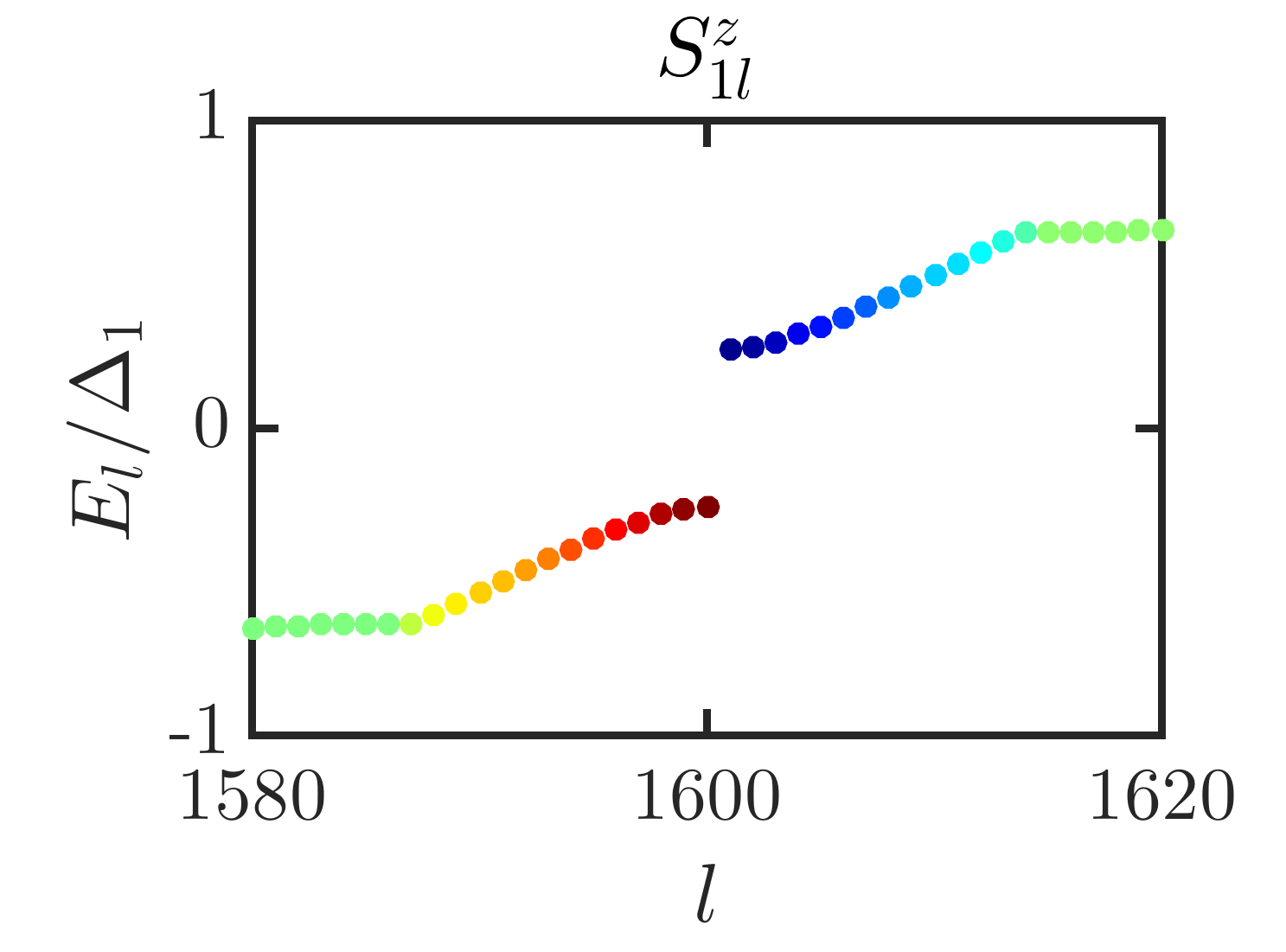,width=1.85in,height=1.45in,clip=true} &\hspace*{-0.4cm}
\epsfig{figure=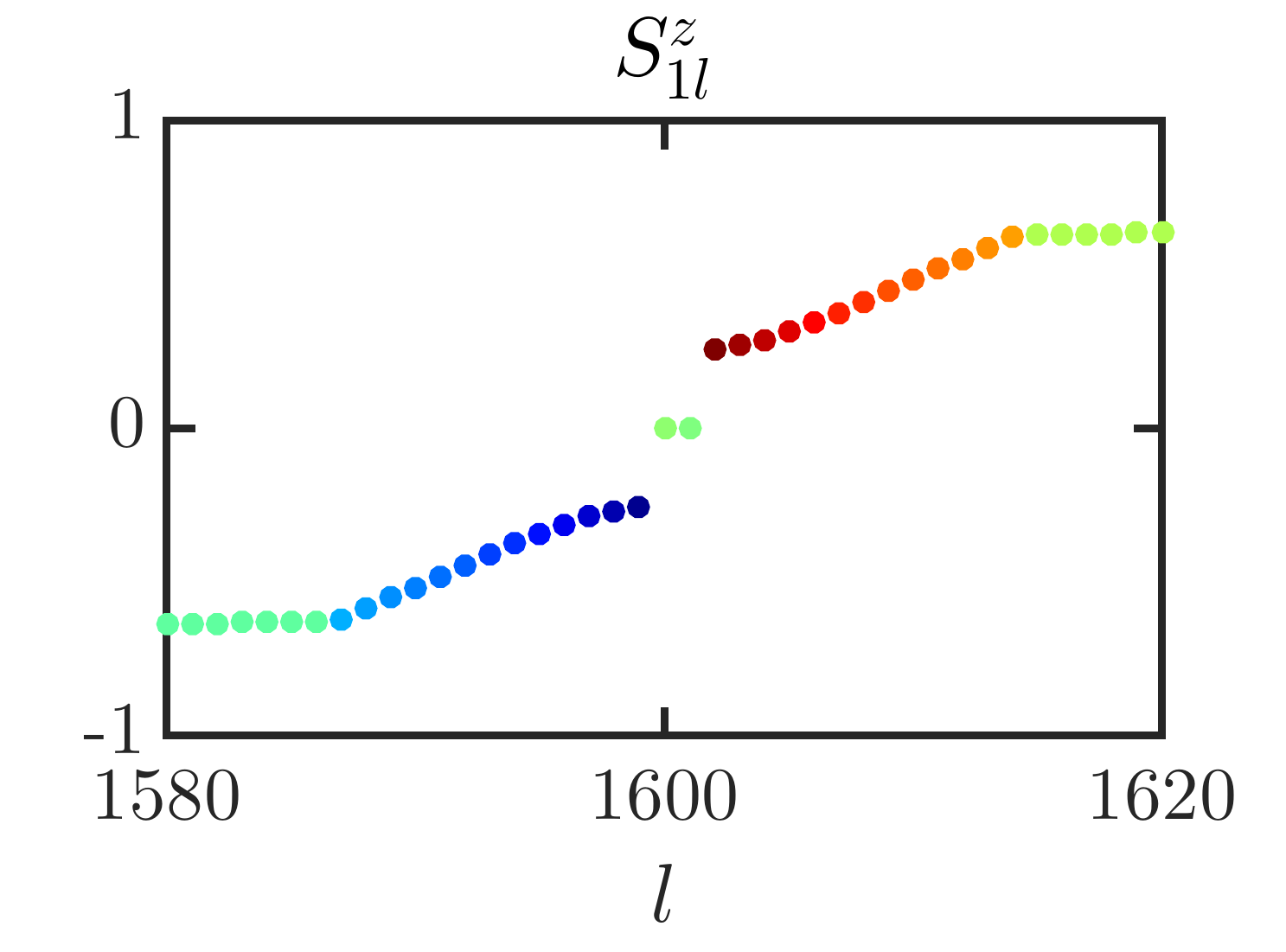,width=1.75in,height=1.45in,clip=true} &\hspace*{-0.4cm}
\epsfig{figure=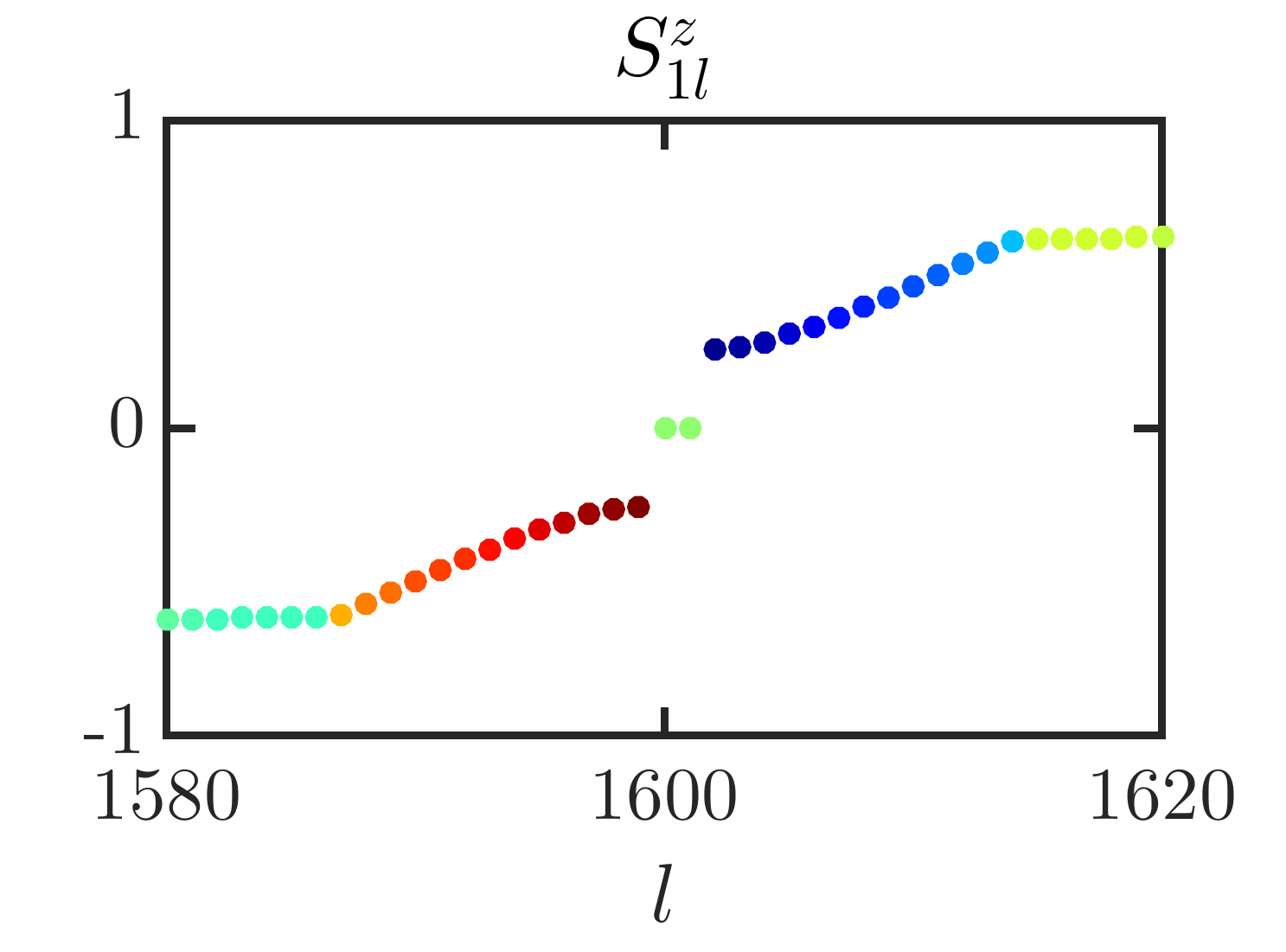,width=1.75in,height=1.45in,clip=true} &\hspace*{-0.4cm}
\epsfig{figure=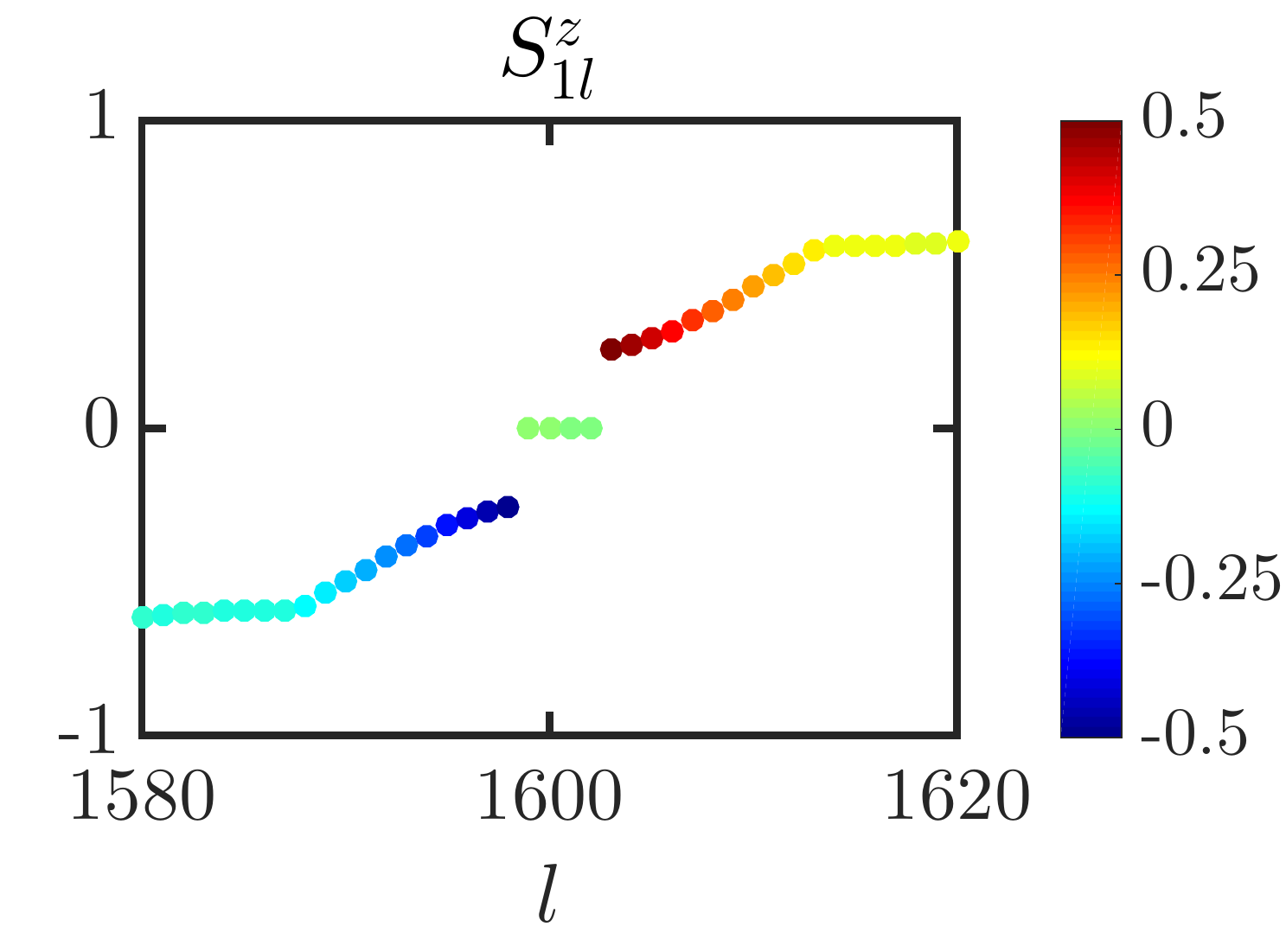,width=1.85in,height=1.45in,clip=true} \\
\epsfig{figure=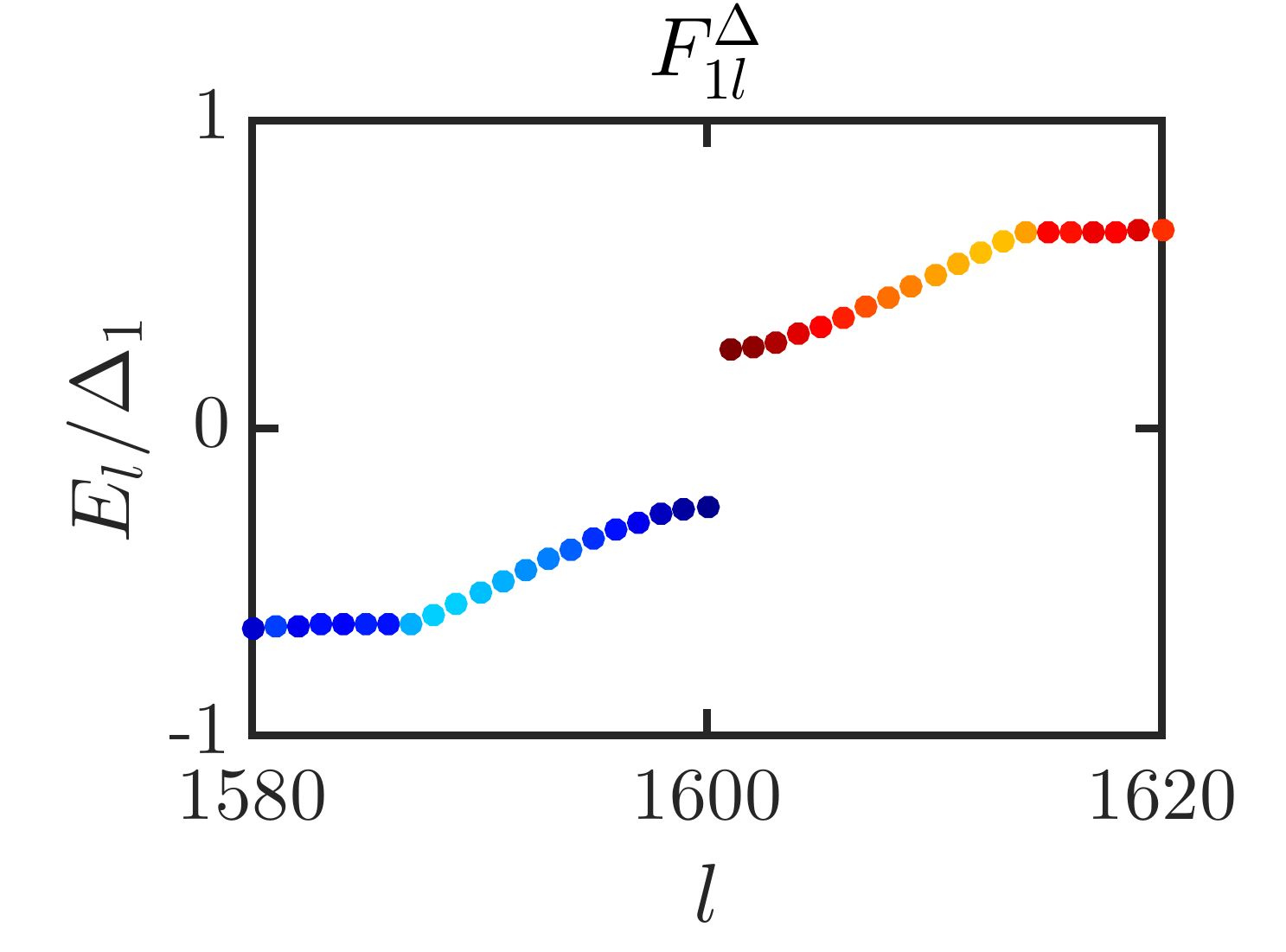,width=1.85in,height=1.45in,clip=true} &\hspace*{-0.4cm}
\epsfig{figure=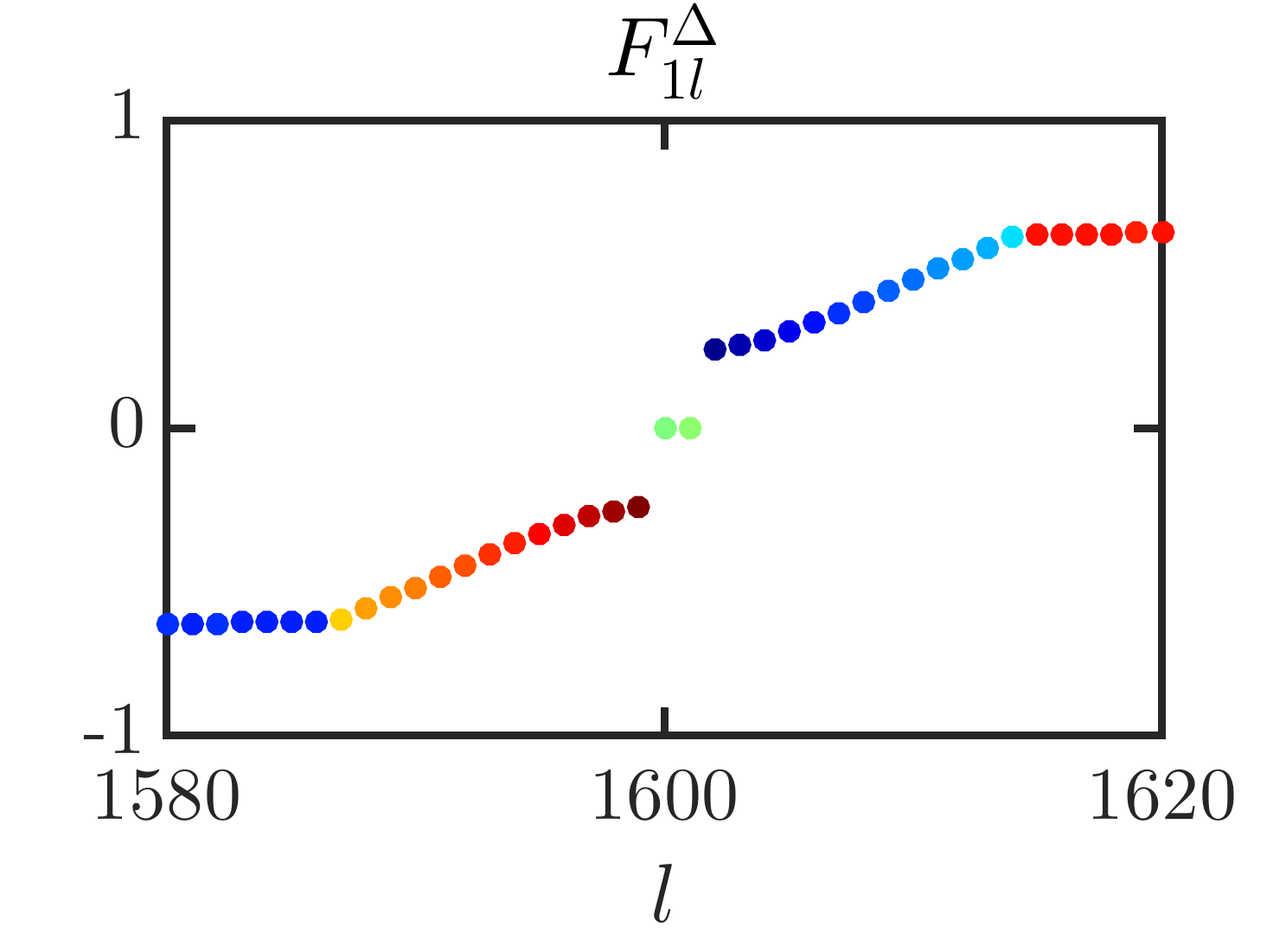,width=1.75in,height=1.45in,clip=true} &\hspace*{-0.4cm}
\epsfig{figure=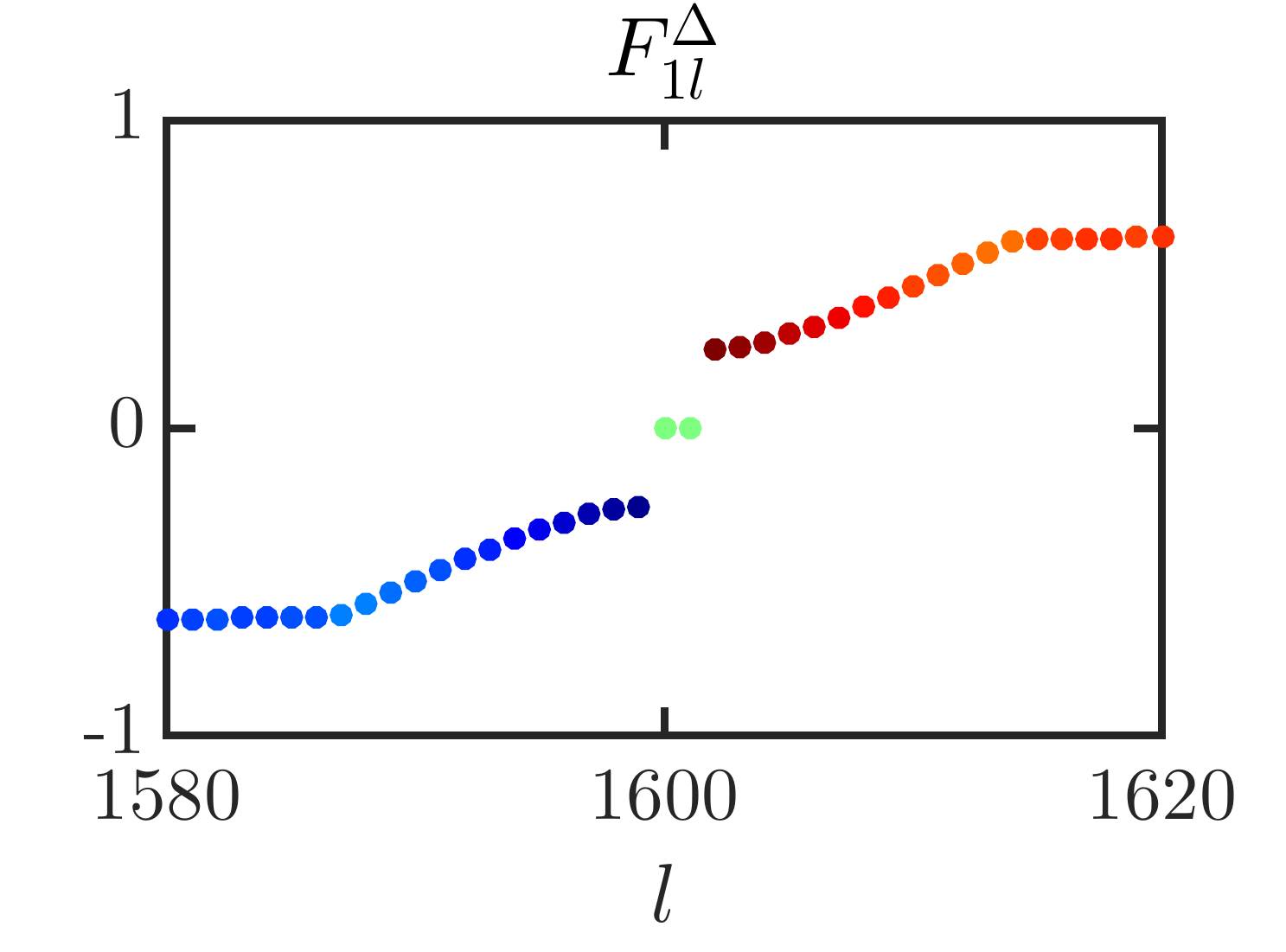,width=1.75in,height=1.45in,clip=true} &\hspace*{-0.4cm}
\epsfig{figure=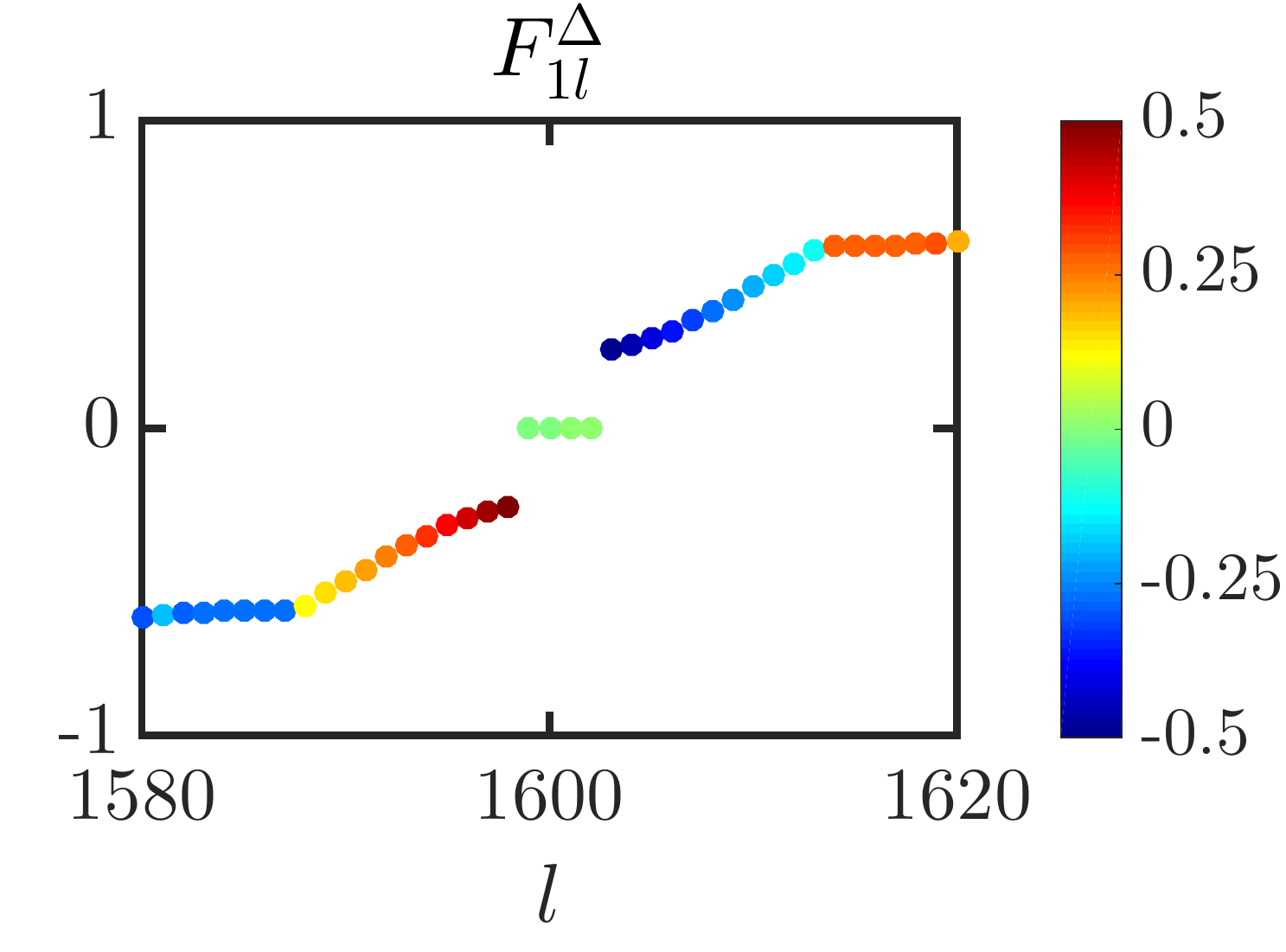,width=1.85in,height=1.45in,clip=true}
\end{tabular} \end{center}
\caption{Energy spectra of the finite double-NW system with open boundary conditions [see Eq. (\ref{H0t})]. The red and blue colors correspond to positive and negative values of the global bulk quantities, the total charge $Q_{\eta l}$, the total spin $z$-component $S^z_{\eta l}$, and for the total intrawire pairing amplitude $F^\Delta_{\eta l} $ in row one, two, and three, respectively. We plot these quantities for NW-1 at four different points $n_1,..., n_4$ in the phase diagram as we go from left to right in each row. The sign flip of  $S^z_{\eta l}$ and $F^\Delta_{\eta l}$ exactly matches with the  sign flip obtained previously for the periodic counterparts $S^z_{\eta \lambda}$ and $F^\Delta_{\eta \lambda}$ shown in Fig. \ref{fig04}. For the numerical simulations, we use $N=800$ sites in each NW and the remaining parameters are the same as in Fig. \ref{fig04}. }
\label{fig05}
\end{figure*}

\section{Signatures of the topological phase transition in charge, spin, and pairing amplitude}
\label{bulk_quant}

In this section we consider equilibrium properties of the double-NW setup, in particular the charge and spin densities as well as the 
intrawire pairing amplitude density in a given eigenstate, and study their behavior as function of momentum (position) and topological phase. We consider 
systems with open  and with periodic boundary conditions. In the latter case, there are no MBSs. Our goal is to find signatures of the topological phase transition in these
quantities.
The bulk densities of interest are then defined as
\begin{align}
&Q_{\eta \lambda}(k)=\Phi^\dagger_{\eta \lambda}(k)\, \tau_z\, \Phi_{\eta \lambda}(k),\label{Qc}\\
&Q_{\eta l}(j)=\Phi^\dagger_{\eta l}(j)\, \tau_z\, \Phi_{\eta l}(j),\label{Ql}\\
&{\bf S}_{\eta\lambda}(k)=\Phi^\dagger_{\eta \lambda}(k)\, {\bm \sigma}\, \Phi_{\eta\lambda}(k),\label{Sc}\\
&{\bf S}_{\eta l}(j)=\Phi^\dagger_{\eta l}(j) { \bm \sigma}\, \Phi_{\eta l}(j),\label{Sl}\\
&F^\Delta_{\eta\lambda}(k)=\Phi^\dagger_{\eta\lambda}(k)\, \tau_x\, \Phi_{\eta\lambda}(k),\label{dc}\\
&F^\Delta_{\eta l}(j)=\Phi^\dagger_{\eta l}(j)\, \tau_x\, \Phi_{\eta l}(j ).\label{dl}
\end{align}
Here, $Q_{\eta \lambda}$ $ (Q_{\eta l})$, $S_{\eta \lambda}$ $(S_{\eta l})$, and  $F^\Delta_{\eta\lambda} $ $(F^\Delta_{\eta l})$ are the densities of the charge, spin, and  intrawire pairing amplitude, respectively, for the $\eta$-NW  at the given energy $E_\lambda$ $(E_l)$ labeled by the index $\lambda$ $ (l)$  found in the continuum (tight-binding) model.  We measure the charge, the spin, and the intrawire pairing amplitude in the units of electronic charge $e$, $\hbar/2$, and $\Delta_1$, respectively. 

To begin with, we consider the setup with periodic boundary conditions that allows us to introduce the momentum $k$ as a good quantum number and to study the bulk quantities as a function of $k$, see Fig. \ref{fig04}.  We follow the line connecting the points $n_1,...,n_4$  shown in Fig. \ref{fig02}. The charge $Q_{\eta \lambda}$ of the lowest energy level  [see Fig. \ref{fig04}]  does not show any prominent sign flip close to $k=0$ as one crosses a topological phase transition line. The same is true for the spin components $S^{x/y}_{\eta \lambda}$.  In contrast to that, the spin component along the magnetic field $S^{z}_{\eta \lambda}$ and the intrawire pairing amplitude $F^\Delta_{\eta\lambda}$ flip their sign as we go along the line $n_1-n_2$ or $n_3-n_4$, indicating the topological phase transition from trivial to topological phase. 

The same results can be obtained analytically by calculating the spin polarization at $k=0$ for the energy levels $E_1$ and $E_2$. Using Eq. (\ref{Sc}), we find that the spin polarization along the direction of the magnetic field for NW-1 in the momentum space is given by $S^z_{11}(k=0) =\text{sgn} (\Delta_{Z1}-\Delta_1+\Delta_c)/2$ for the level $E_1$ and by $S^z_{12}(k=0) =\text{sgn} (\Delta_{Z1}-\Delta_1-\Delta_c)/2$ for the level $E_2$. This explains the spin flip when moving along the line $n_1-n_2$ ($n_3-n_4$), where $E_1$ ($E_2$) is the lowest energy level.  Another spin flip occurs between $n_2$ and $n_3$. However, this flip does not correspond to a topological phase transition but instead corresponds to the reordering of the two lowest bands  $E_1$ and $E_2$ (which have opposite values of spin and  pairing amplitude) at the point $\Delta_{Z1}= \Delta_1$. In other words, when magnetic field is tuned to $\Delta_{Z1}= \Delta_1$,  the energy bands $E_1$ and $E_2$ are degenerate at $k=0$. As a result, we observe the sign flip in the spin and pairing amplitude at this point. Due to the strong spin polarization around $k=0$, such spin flips can be accessed in transport experiments
as we show in the next section. We further note that the spin flip can also be accessed locally as the spin density of the lowest energy band at $k=0$ is uniform along the NWs. Due to the translation invariance, the signal can be measured  at any point of the NW as long as it is sufficiently far away from the NW ends.
 
To show the agreement between  the continuum and the tight-binding model, we obtain the bulk quantities for the finite system with open boundary conditions by numerically solving the tight-binding Hamiltonian defined in Eq. (\ref{H0t}). A similar behaviour [see Fig. \ref{fig05}] is found for the global (total)  bulk quantities defined as $Q_{\eta l}=\sum_{j=1}^NQ_{\eta l}(j)$, ${\bf S}_{\eta l}=\sum_{j=1}^N {\bf S}_{\eta l}(j)$, and $F_{\eta l}^\Delta=\sum_{j=1}^NF^\Delta_{\eta l}(j)$. In the next section, we calculate the spin current probed by a weakly coupled spin-polarized STM tip. The tip is sensitive to the local spin polarization in the given band. However, due to the translation invariance of the setup, one can argue that the sign flip occurs for  both, the local $ S^z_{\eta l}(j)$ and the global spin component  $S^z_{\eta l}$. 
 
 \begin{figure*}[t]
\begin{center}\begin{tabular}{cccc}
\epsfig{figure=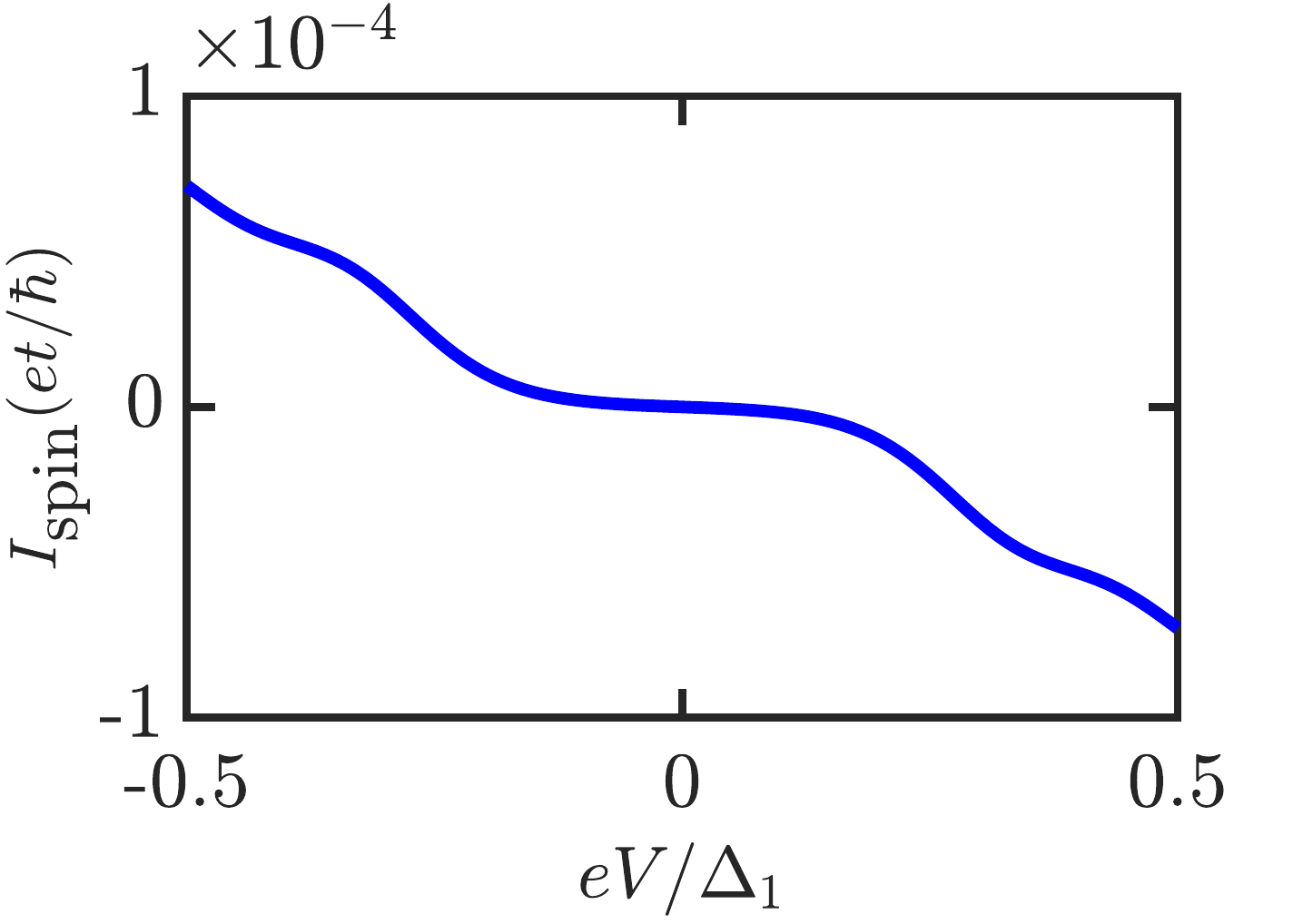,width=1.85in,height=1.38in,clip=true} &\hspace*{-0.4cm}
\raisebox{-0.01cm}
{\epsfig{figure=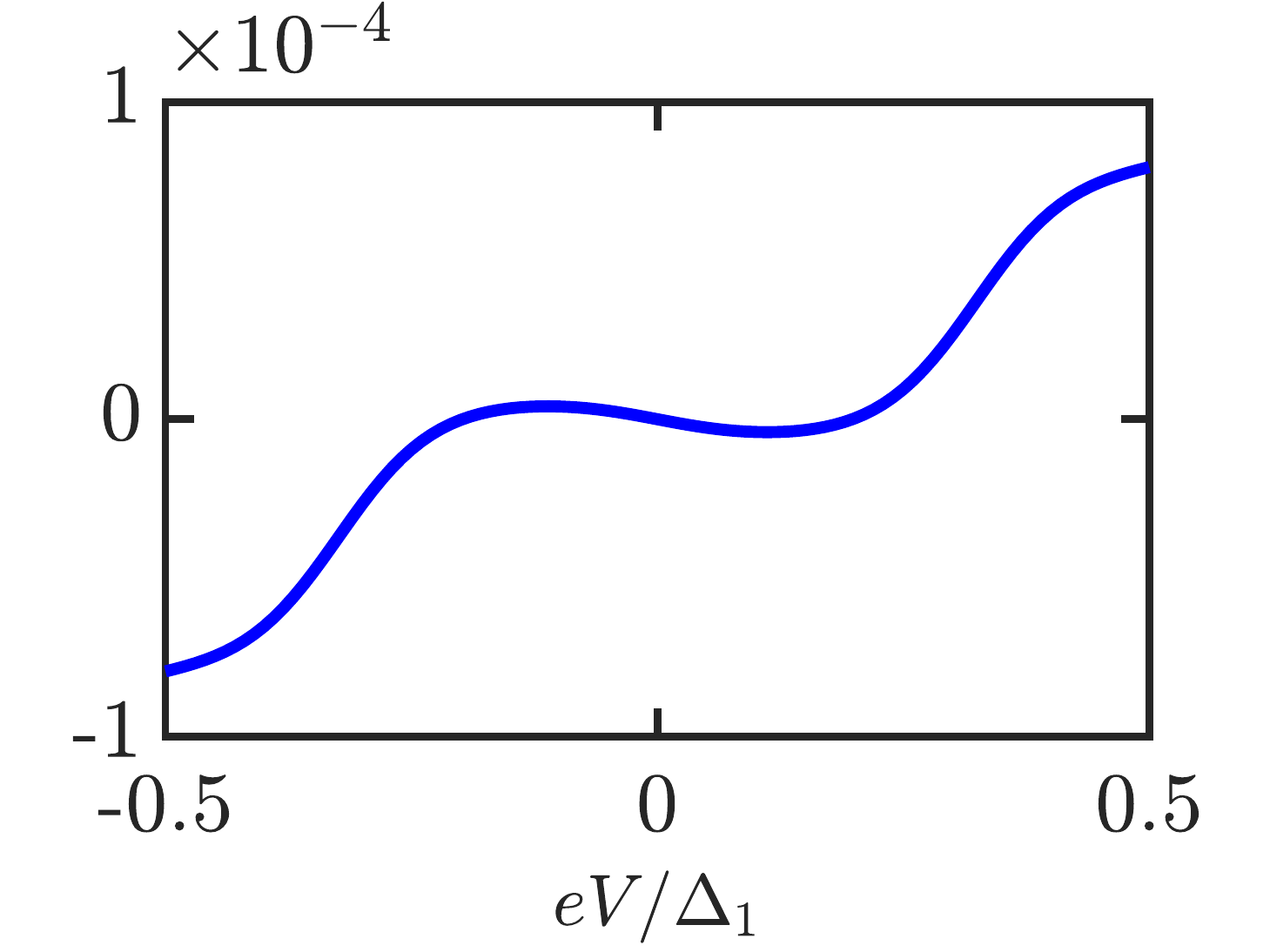,width=1.75in,height=1.38in,clip=true}}
&\hspace*{-0.4cm}
\raisebox{-0.0cm}
{\epsfig{figure=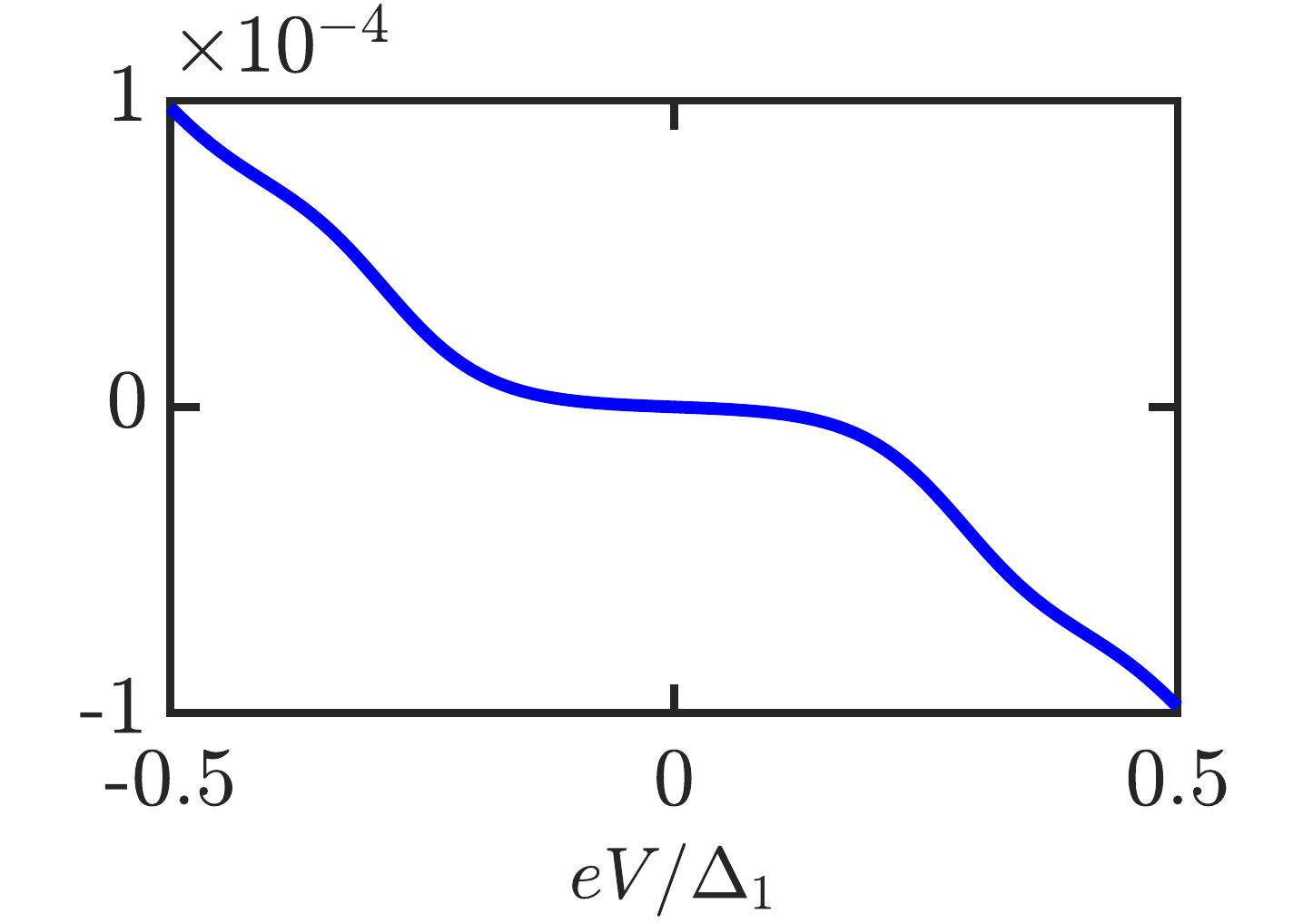,width=1.75in,height=1.38in,clip=true}} &\hspace*{-0.4cm}
\raisebox{-0.03cm}
{\epsfig{figure=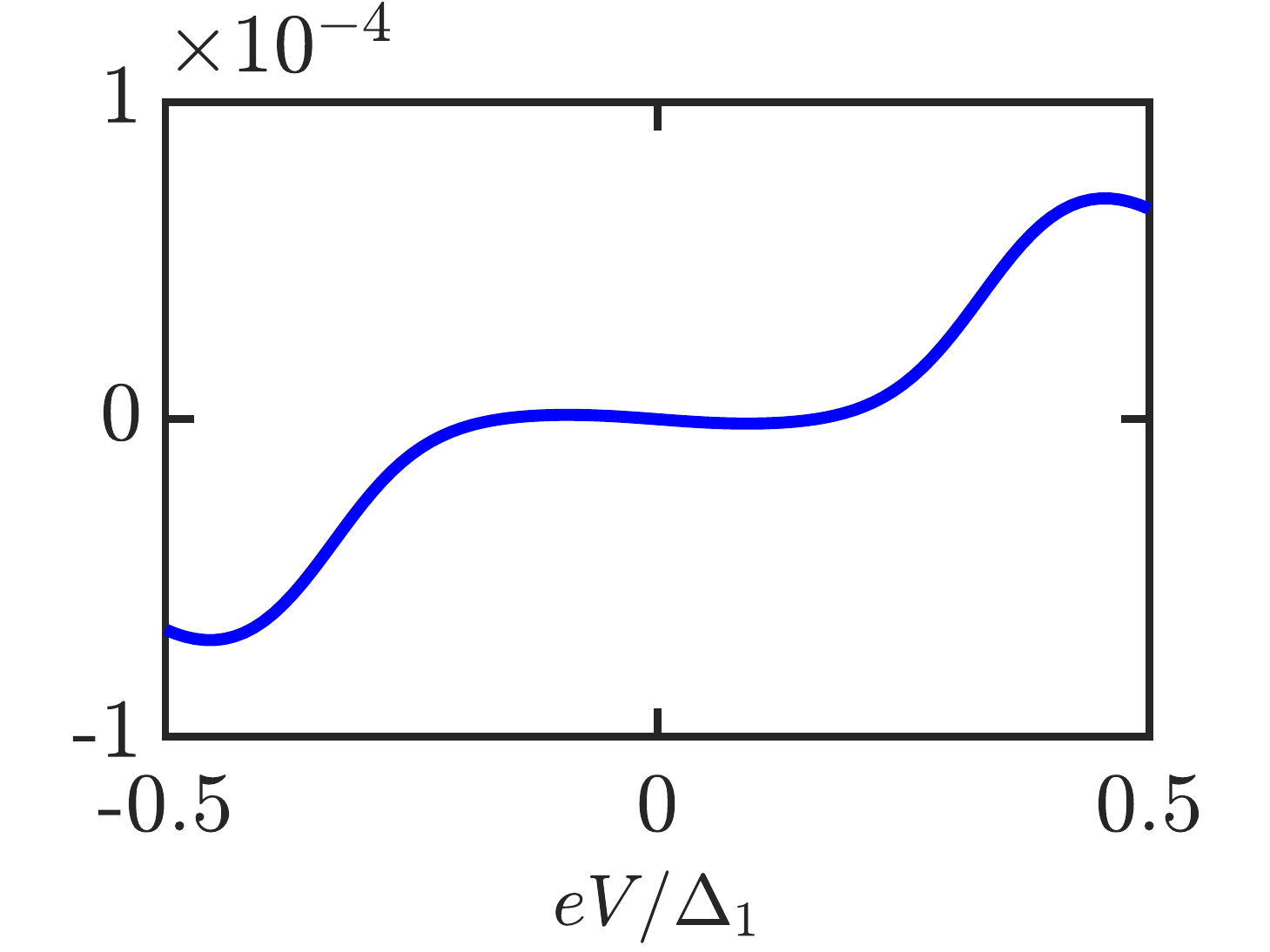,width=1.75in,height=1.38in,clip=true}}\\
\epsfig{figure=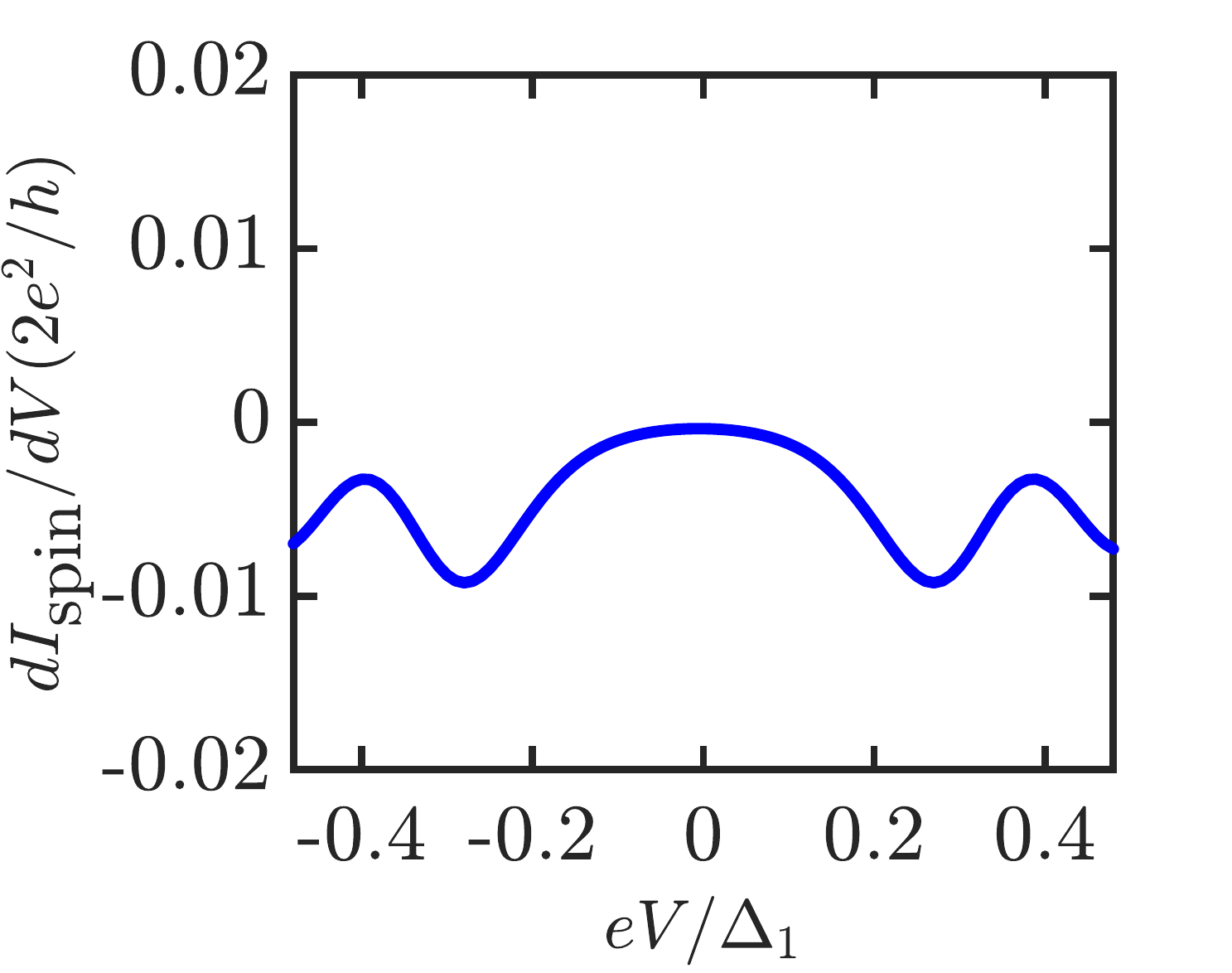,width=1.85in,height=1.38in,clip=true} &\hspace*{-0.4cm}
{\epsfig{figure=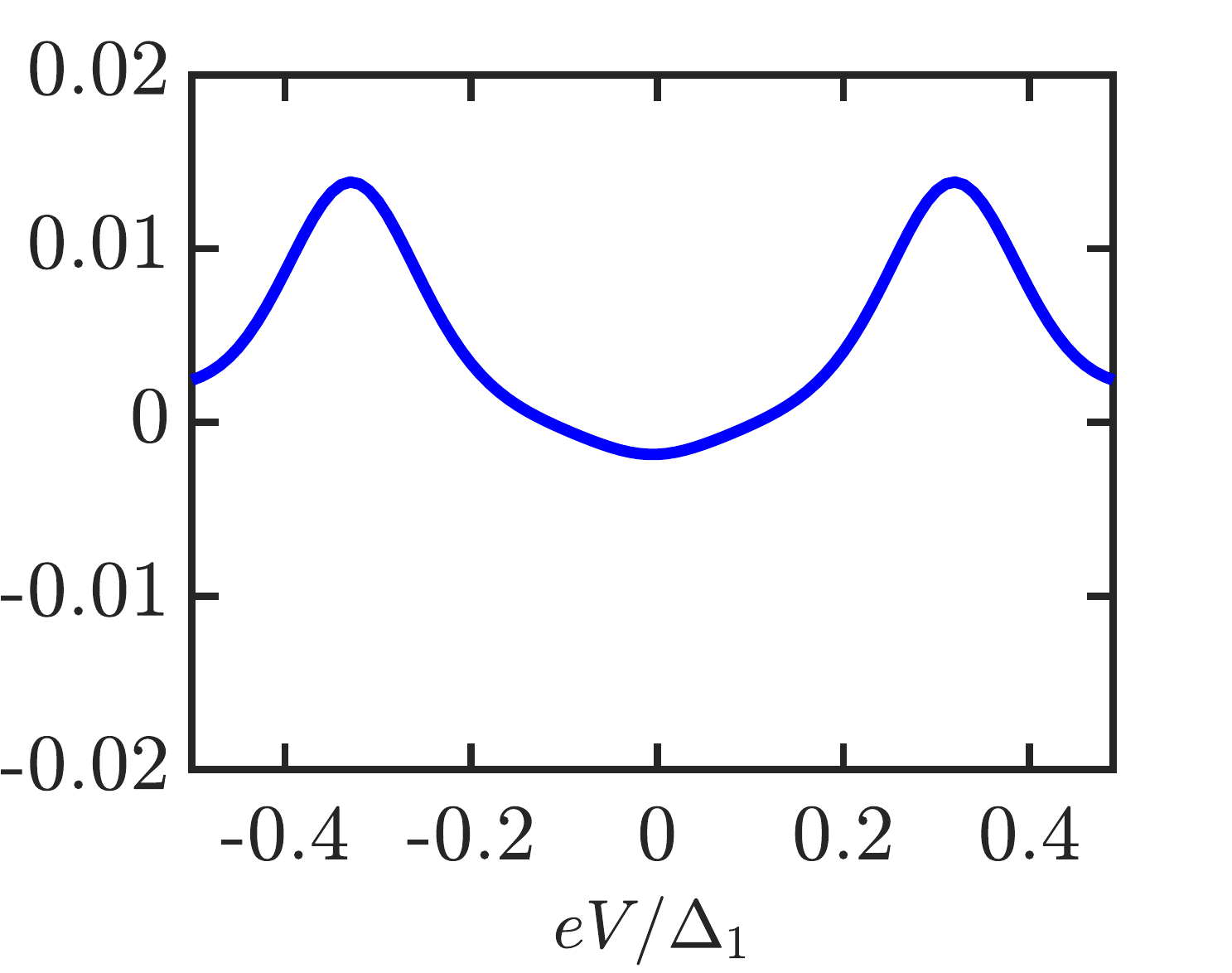,width=1.75in,height=1.38in,clip=true}}
&\hspace*{-0.4cm}
{\epsfig{figure=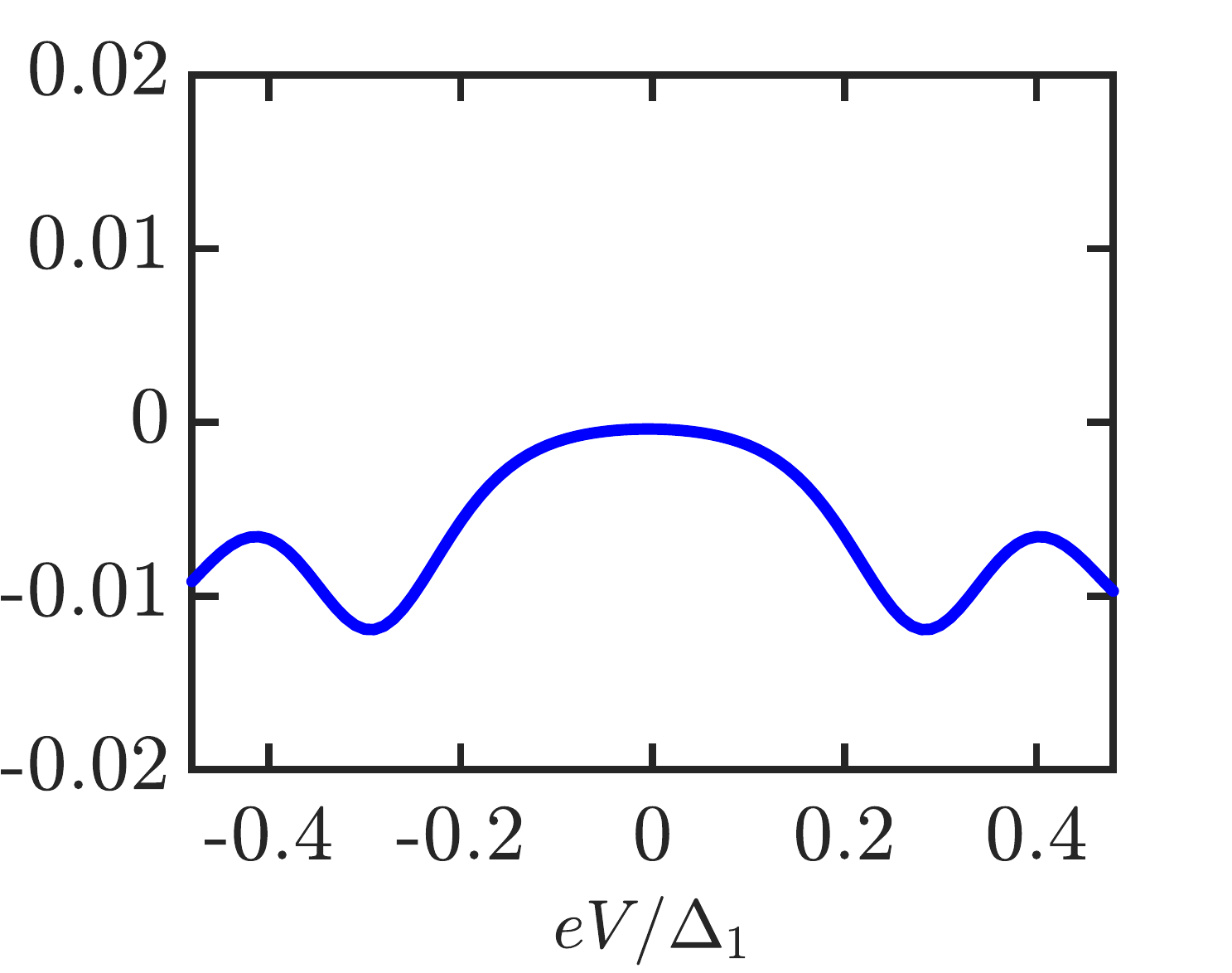,width=1.75in,height=1.38in,clip=true}} &\hspace*{-0.4cm}
{\epsfig{figure=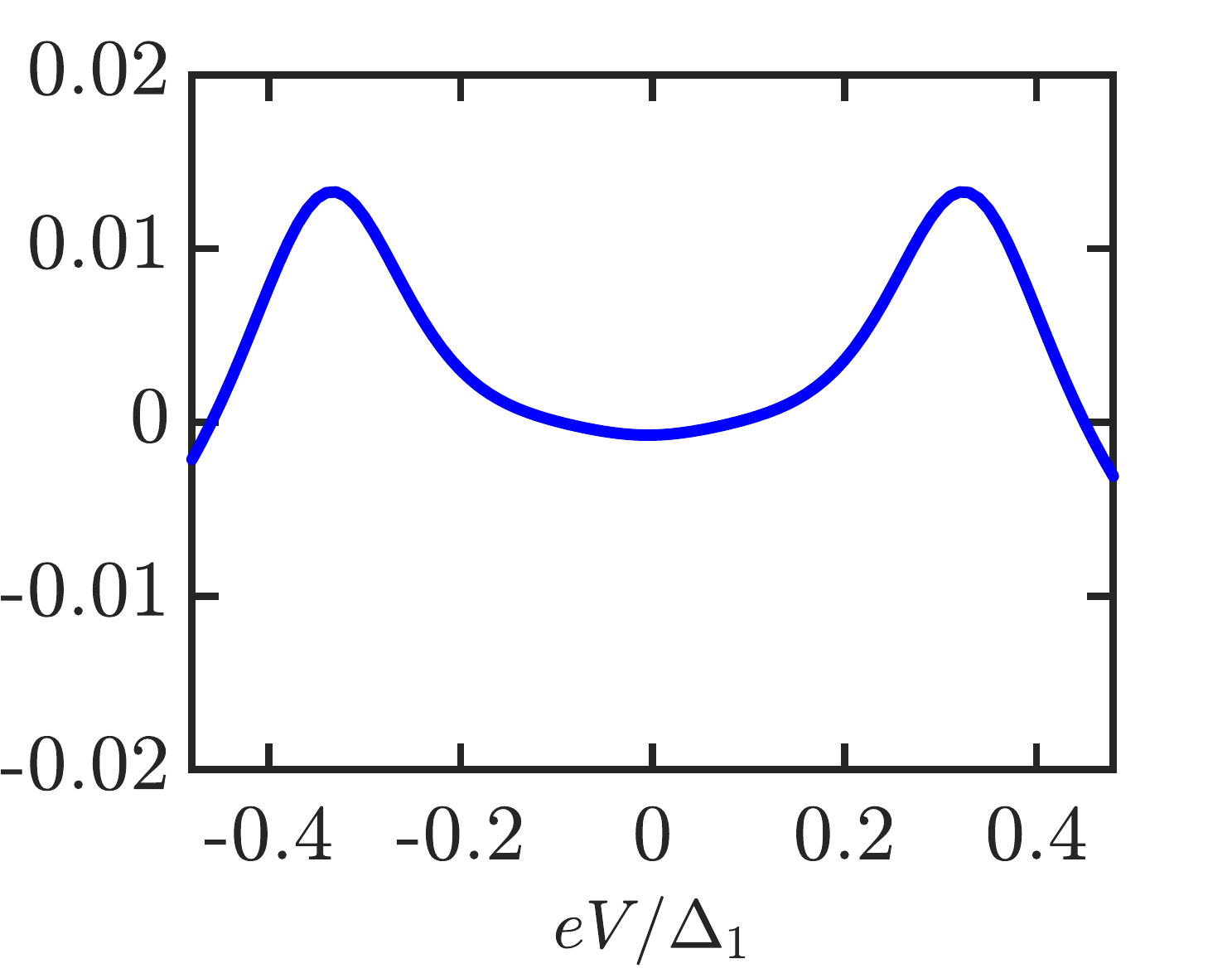,width=1.75in,height=1.38in,clip=true}}
\end{tabular}
\end{center}
\caption{Plots of the spin current and the differential conductance as a function of bias voltage. In the first row,  we plot $I_{\text{spin}}$ flowing between the SP STM tip and the first NW  for four different points $n_1-n_4$ in the topological phase diagram as we go from left to right. To compute the bulk contribution, the SP STM tip is placed in the middle of the NW, {\it i.e.} at site $j=N/2$.  In the second row, we plot the corresponding differential conductance $d I_{\text{spin}}/dV$ from the spin current. The differential conductance also flips its sign as one goes along along the line $n_1-n_4$ from left to right indicating the topological phase transition. Here, only the lowest band is probed, $|E_l/\Delta_1|<0.5$ ($|E_{\lambda}/\Delta_1|<0.5$), defining the range of the applied bias voltage $eV/\Delta_1=[-0.5,0.5]$. The spin current and its differential conductance clearly captures the sign-flip behaviour of the $z$ component of the spin, compare with Figs. \ref{fig04} and \ref{fig05} . Here, we take $N=100$ sites for each NW, $\Gamma_t/\Delta_1=0.1$, $k_BT/\Delta_1=1/20$, and other parameters are the same as in Figs. \ref{fig04} and \ref{fig05}.}
\label{fig06}
\end{figure*}

\section{Calculation of the spin current probed by a spin-polarized STM}
\label{SP}

To model transport measurements probed by a spin-polarized STM, we calculate the spin current using the Keldysh formalism. The measurement setup consists of two parts, namely  of the lead formed by the SP STM tip and of the substrate (double-NW setup). We further introduce the Hamiltonian corresponding to the lead,
\begin{align}
H_l= \sum_k \Psi^\dagger_k \,\xi_k \,\tau_z\, \Psi_k,
\label{Hl}
\end{align}
where the components of the spinor $\Psi_k=(\psi^\dagger_{k, \uparrow}, \psi^\dagger_{k,\downarrow}, \psi_{k,\downarrow}, -\psi_{k,\uparrow})$ written  in the Nambu basis correspond to the operators acting on electrons inside the SP STM tip. Here, $\xi_k= \hbar^2 k^2/2m -\mu$ is the energy dispersion relation with $m$ being the effective mass. The tunneling Hamiltonian between the lead and the substrate has following form
\begin{align}
H_T(t)=\sum_k \Psi_k^{\dagger}\mathcal T( t)\,\Phi_{\eta j} +\text {H.c.}
\label{Ht}
\end{align}
Here, we have included the voltage bias $V$ in the tunneling amplitude $\mathcal{T}(\bar t)=\bar t_j\,\tau_z e^{i\tau_z\sigma_z V\,t}$, where $\bar t_j$ is the tunneling amplitude between the tip and the site $j$ of the double-NWsetup. We remind the reader that $\Phi_{\eta j}$ is the electronic operator acting in the $\eta$-NW.  The retarded and advanced Green functions for the SP STM tip have the form [\onlinecite{DC}]
\begin{align}
g_{R/A}^s(\omega)=\int d\xi \,\nu^s(\xi)\, [(\omega \pm i \gamma)- \xi\, \tau_z]^{-1}.
\end{align}
The spin-dependent density of state is written as $\nu^s(\xi)=(1-P^s \sigma_z)(1-\tau_z)\nu_0/4$ with $\nu_0=\sum_k \delta(\xi-\xi_k)$, where $\nu_0$ is  assumed to be constant at the Fermi energy.  Here, $P^s=1(\bar 1)$ corresponds to the spin polarization of the electron state in the SP STM tip along (opposite to) the applied magnetic field. In principle, one can consider $|P^s| \leq 1$, however, to obtain the maximum spin current, we assume a fully 
spin-polarized STM tip with $|P^s|= 1$. As a result, the Green function becomes independent of the frequency $\omega$,
\begin{align}
g_{R/A}^{s}(\omega)&=\frac{\nu_0}{4}\,(1-P^s \sigma_z)(1-\tau_z)\nn
&\times \Big[\omega \int \frac{d\xi}{(\omega\pm i\gamma)^2-\xi^2}+\int \frac{\xi \tau_z d\xi}{(\omega\pm i\gamma)^2-\xi^2}\Big]\nn
&= \mp i\,\frac{\pi\,\nu_0} {4}(1-P^s \sigma_z)(1-\tau_z).
\end{align}

We also write the Keldysh Green function for the SP STM tip as
\begin{align}
g_K^{s}(\omega)=[1-2f(\omega)][g_R^{s}(\omega)-g_A^{s}(\omega)],
\end{align}
where $f(\omega)=1/(1+e^{\beta\omega})$ with inverse temperature $\beta=1/k_BT$. The on-site self-energy of the SP STM  [for details, see App. \ref{App_SE}] has the following form:
\begin{align}
&\Sigma_{R/A}^{s,j}=\mp\, i\, \Gamma^{s,j} ,\\
&\Sigma_{K}^{s,j}=-2\, i\, \Gamma^{s,j}\nn
 \times&\begin{pmatrix}
\text{tanh}\frac{\beta(\omega-V)}{2}& 0&0&0\\
0& \text{tanh}\frac{\beta(\omega-V)}{2}&0&0\\
0&0& \text{tanh}\frac{\beta(\omega+V)}{2}&0\\
0&0&0& \text{tanh}\frac{\beta(\omega+V)}{2}
\end{pmatrix},\nonumber
\end{align}
where the tunneling rate $\Gamma^{s,j}=\Gamma_{\bar t} (1-P^s \sigma_z)(1-\tau_z)$ is the spin-dependent coupling strength with $\Gamma_{\bar t}= \pi\nu_0|t_{j}|^2/4$. The full form of the retarded or advanced self-energy of the SP STM tip can be written in  real space as $[\Sigma^s_{R/A}(\omega)]_{jj}=\Sigma^{s,j}_{R/A}(\omega)$ with the only  non-zero component at the site at which the tip is connected to the double-NW setup. For our calculation, without loss of generality, we choose to work at the middle of the NW, $j=N/2$. We note that the signal coming from the MBSs can spoil the
desired current contribution coming from the bulk. Therefore, to avoid this issue, the SP STM tip should be placed sufficiently  far away from  the NW ends such that the contribution from the zero-energy states (MBSs) is negligible. Next, one can obtain the total Green function using the following relations:
\begin{align}
&[G^s_{R/A}(\omega)]^{-1}= G_{0R/A}^{-1}-\Sigma^s_{R/A}(\omega),\nn
&G^s_K(\omega)=G_{0K}+G^s_R(\omega)\Sigma^s_K(\omega)G^s_A(\omega).
\end{align}
The Green function $G_{0K}$ is zero in the rotated Keldysh basis [\onlinecite{DC}]. Further, we utilize the Green function and the Keldysh technique, as discussed in Apps. \ref{App_SE} and \ref{App_CC}, and obtain the expression for the current, 
\begin{align}
I_{DC}^{s}=\frac{e}{\hbar}Tr\Bigg( \tau_z \int_{-\infty}^{\infty} &\frac{d\omega}{2\pi} Re[G^s_R(\omega)\Sigma^{s}_K(\omega)\nn
&\hspace{40pt}+G^s_K(\omega)\Sigma^{s}_A(\omega)]\Bigg).
\end{align}
The spin-filtered current, or spin current for short, $I_\text{spin}$ is the difference of the spin-up $I_{DC}^{+}$ and spin-down $I_{DC}^{-}$ currents, where $+ (-)$ corresponds to fully spin-polarized SP STM tip with $P_s=1 (-1)$, see Fig.   \ref{fig06}.
The stronger the SP STM tip couples to the substrate, the larger is the spin current $I_\text{spin}$.
The pattern of $I_\text{spin}$, arising as one goes along the line $n_1-n_4$ in the topological phase diagram, verifies the sign flip of the spin component  along the magnetic field at the topological phase transition points as was shown in Figs. \ref{fig04} and \ref{fig05}. One can also see from Fig. \ref{fig06} that the differential conductance of the spin current $dI_\text{spin}/dV$ also flips its sign as the system transitions from trivial to topological phase. In the trivial phase ($n_1$),  the spin current decreases, however, after the first topological phase transition ($n_2$), the spin current increases as a function of the voltage bias. Moreover, if we connect the tip at the end of the NW, the non-zero contribution of the MBS appears at zero bias, see Fig. \ref{fig07} in App. \ref{App_CC}. 
However, for the STM  measurement, all the energy levels below the Fermi level contribute to the spin current and as a result signals coming from the MBSs mask the bulk contribution. Therefore, a clear signature of the sign flip of the bulk spin current and differential conductance does not emerge in this case, which emphasizes the importance of probing the bulk properties of the system sufficiently far away from the NW ends, meaning at a distance which exceeds the localization length of the MBSs.

\section{Conclusions} \label{con}
In this work, we studied a double-NW setup proximity coupled to an $s$-wave superconductor in the presence of Rashba SOI and subjected to a magnetic field along the NW. This setup has a richer phase diagram compared to a single NW setup because of the competition between three  gap opening mechanisms, namely, the intrawire and the interwire superconductimg pairings as well as  the magnetic field. We analyzed three physical bulk densities of charge, spin, and intrawire pairing amplitude, which can be experimentally observed. The latter two flip sign as the system goes from the trivial to the topological phase. To detect this sign flip experimentally, we propose to perform transport measurements with the use of a weakly coupled spin-polarized STM. Using the Keldysh technique, we demonstrated that the spin current through a weakly coupled STM which filters the spin component along the direction of the applied magnetic field, fully captures the sign flip of the spin due to the topological phase transition. 
These findings show that  spin-polarized local transport probes, such as STMs, provide a powerful tool to detect experimentally topological phase transitions.
This type of bulk measurement constitutes an alternative approach to detect topological superconductivity that avoids the ambiguity associated with the zero-bias peak coming from zero-energy bound states located at the end of the nanowires.
 
\section*{Acknowledgments}
We acknowledge support  from the Swiss National Science Foundation and NCCR QSIT. This project received funding from the European Union Horizon 2020 research and innovation program (ERC starting grant, Grant Agreement No. 757725).

\appendix
\onecolumngrid
\section{Self-energy of the SP STM tip}
\label{App_SE}
In this appendix, we compute the retarded, advanced, and Keldysh part of the on-site self-energy for the SP STM tip that enters via the tunneling term and is given by [\onlinecite{DC}]
\begin{align}
\Sigma^{s,j}(t_1,t_2)= \mathcal{T}^\dagger(t_1) \pi_z g^s(t_1-t_2)\pi_z \mathcal{T}(t_2).
\label{SE1}
\end{align}
Here, the Pauli matrix $\pi_z$ acts in Keldysh space. The tunneling amplitude includes the voltage dependence, $\mathcal{T}( t)=\bar t_j\,\tau_z e^{i\tau_z V\,t}$, and $g^s= L^\dagger \tilde g^s L$, where $L$ is a unitary transformation rotating the Keldysh basis, 
\begin{align}
\tilde g^s= \frac{1}{\sqrt 2} \begin{bmatrix}
0 & g_A^s\\
g_R^s & g_K^s
\end{bmatrix}
\text{and}~L= \frac{1}{\sqrt 2} \begin{bmatrix}
1 & -1\\
1 & 1
\end{bmatrix}.
\end{align}
We write the Green function for the SP STM tip in the rotated Keldysh basis as
\begin{align}
\pi_z g^s(t_1-t_2)\pi_z=\pi_z L^\dagger \tilde g^s(t_1-t_2)L \pi_z=\int \frac{d\omega}{2\pi} e^{-i\omega'(t_1-t_2)} \frac{1}{2}\begin{pmatrix}
1 &1\\1&-1
\end{pmatrix} \begin{pmatrix}
0 & g^s_A(\omega')\\
g^s_R(\omega') & (1-2f_{\omega'}) [g^s_R(\omega')-g^s_A(\omega')]
\end{pmatrix}\begin{pmatrix}
1 &1\\1&-1
\end{pmatrix}.
\end{align}
Therefore, the self-energy in Eq. (\ref{SE1}) takes the form
\begin{align}
\Sigma^{s,j}(t_1,t_2)=i\,\Gamma^{s,j}\,\int_{-\infty}^{\infty} \frac{d\omega'}{2\pi} e^{-i\omega'(t_1-t_2)} e^{-i\tau_z V t_1}\mathds{1} e^{i\tau_z Vt_2}\begin{pmatrix}
2f_{\omega'}-1 & -2f_{\omega'}\\
2-2f_{\omega'} & 2f_{\omega'}-1
\end{pmatrix},
\end{align} 
where, for convenience, we use the notation $\Gamma^{s,j}=\Gamma_{\bar t}(1-P^s \sigma_z)(1-\tau_z)$ with $\Gamma_{\bar t}=\pi\nu(0)|\bar t_j|^2/4$. The Fourier transform of the self-energy is given by
\begin{align}
\Sigma_{nm}^{s,j}(\omega)&=\frac{1}{2\pi}\int_{-\infty}^{+\infty}\int_{-\infty}^{+\infty} dt_1\, dt_2\, e^{i(\omega+nV)t_1}e^{-i(\omega+mV)t_2}\Sigma^s(t_1,t_2)\nn
&= \frac{\Gamma^{s,j}}{4\pi^2}\int_{-\infty}^{+\infty}\int_{-\infty}^{+\infty} dt_1\, dt_2\, e^{i(\omega+nV)t_1}e^{-i(\omega+mV)t_2}\int_{-\infty}^{+\infty} d\omega'e^{-i(\omega'+\tau_zV)(t_1-t_2)} \begin{pmatrix}
2f_{\omega'}-1 & -2f_{\omega'}\\
2-2f_{\omega'} & 2f_{\omega'}-1
\end{pmatrix}\nn
&=i\,\Gamma^{s,j} \int d\omega'
\delta(\omega+nV-\omega'-\tau_zV)\delta(\omega+mV-\omega'-\tau_z V)\begin{pmatrix}
2f_{\omega'}-1 & -2f_{\omega'}\\
2-2f_{\omega'} & 2f_{\omega'}-1
\end{pmatrix}\nn
&=i\,\Gamma^{s,j} \begin{bmatrix}
\delta_{n,m} X(\omega+nV-V)&0&0&0\nn
0& \delta_{n,m} X(\omega+nV-V)&0&0\nn
0&0& \delta_{n,m} X(\omega+nV+V)&0\nn
0&0&0&\delta_{n,m} X(\omega+nV+V)
\end{bmatrix},
\end{align}
where \begin{align}
X(\omega)=\begin{pmatrix}
2f_{\omega}-1 & -2f_{\omega}\\
2-2f_{\omega} & 2f_{\omega}-1
\end{pmatrix}=\begin{pmatrix}
X^{++} & X^{+-}\\
X^{-+} & X^{--}
\end{pmatrix}.
\end{align}
For stationary currents, we consider $n = m = 0$. Further, we calculate the retarded, advanced, and Keldysh part of $X$ as
\begin{align}
X_A&= X^{++}+X^{-+}=1,\nn
X_R&= X^{++}+X^{+-}=-1,\nn
X_K&=X^{++}+X^{--}=-2\tanh \left(\frac{\beta \omega}{2}\right).
\end{align}
Therefore, the final form of retarded, advanced, and Keldysh parts of the self-energy reads as
\begin{align}
\Sigma^{s,j}_{R/A}(\omega)=\mp i \Gamma^{s,j}, \Sigma^{s,j}_{K}(\omega)=-2 i \Gamma^{s,j}\begin{bmatrix}
\tanh \frac{\beta (\omega-V)}{2}&0&0&0\\
0&\tanh \frac{\beta (\omega-V)}{2}&0&0\\
0&0&\tanh \frac{\beta (\omega+V)}{2}&0\\
0&0&0&\tanh \frac{\beta (\omega+V)}{2}
\end{bmatrix}.
\end{align} 
The total retarded or advanced self-energy has diagonal form with $[\Sigma^s_{R/A}(\omega)]_{jj}=\Sigma^{s,j}_{R/A}(\omega)$ and  the only non-zero component arises at the site at which the SP STM tip is connected to the double-NW setup.

\section{Spin current calculation}
\label{App_CC}
We use the Keldysh formalism to calculate the spin current. We introduce the counting field $\eta$ such that the Keldysh partition function has the form $Z=\text{Tr[exp}({-\beta H_0}) S(\infty,\eta)]$, where $S(\infty,\eta)=T_c\, \text{exp}[-i\int_{-\infty}^{\infty}dt \mathcal{H}_T( t,\eta)]$ with $\mathcal{H}_T(\bar t,\eta)=\sum_k \Psi^\dagger_k \mathcal{T}( t)\pi_ze^{i\pi_z\tau_z \eta(t)} \Phi_{\eta j}(t)$, where $T_c$ is the time-ordering operator along the Keldysh contour $c$.
The spin current is defined as 
\begin{align}
\langle I_s(t)\rangle= \left[\frac{i \,e}{\hbar Z_0}\frac{\partial Z[\eta(t)]}{\partial\eta(t)}\right]_{\eta=0},
\label{I2}
\end{align}
where $H_0=H+H_l$ with $H$ [$H_l$] defined in Eq. (\ref{H}) [Eq. (\ref{Hl})] and $Z_0= \text{Tr [exp}(\beta H_0)]$. 
First, we calculate the partition function as follows
 \begin{align}
 Z&=\text{Tr}[e^{-\beta H_0} T_c e^{-i\int_0^\beta d t \mathcal{H}_T}],
 \end{align}
where $\mathcal{H}_T(t)=\sum_k \Psi^\dagger_k \mathcal{T}( t) \Phi_{\eta j}(t)+$ H.c. Second, we expand the exponential in $Z$ in the  tunneling Hamiltonian and use
  \begin{align}
 \overline{\mathcal{H}_T(t_1)\mathcal{H}_T(t_2)}=2 \,T_c \sum_{k,k'} \Phi_{\eta j}^\dagger(t_1)\mathcal{T}^\dagger(t_1)\overline{\Psi_{k}(t_1) \Psi^\dagger_{k'}(t_2)} \mathcal{T}(t_2)\Phi_{\eta j}(t_2),
  \end{align}
  where $\overline{\Psi_{k}(t_1) \Psi^\dagger_{k'}(t_2)}= \langle T_c[\Psi_k(t_1) \Psi_k'(t_2)]\rangle= i \delta_{k,k'}g^s(t_1-t_2)$. Therefore, the partition function takes the form
  \begin{align}
  Z=\left\langle e^{-\beta H_0} T_c\, \text{exp}\left(-i\int_0^\beta dt_1 dt_2 \,\Phi_{\eta j}^\dagger(t_1)\mathcal{T}^\dagger(t_1) g^s(t_1-t_2)\mathcal{T}(t_2)\Phi_{\eta j}(t_2) \right)\right\rangle_{H}.
\end{align}   
  In the Keldysh-Nambu space, we introduce a coupling field $\eta$ such that $\mathcal{T}(t)\rightarrow \mathcal{T} (t) \pi_ze^{i \pi_z\tau_z \eta(t)/2}$. Therefore, utilizing the definition of the self-energy of the SP STM tip given in Eq. (\ref{SE1}), the partition function becomes  
 \begin{align}
 \frac{Z[\eta]}{Z_0}&= Tr_{\text{Keldysh-Nambu}}\langle\langle\cdots\rangle_{H_l}\rangle_{H}\nn
 &=Tr_{\text{Keldysh-Nambu}}\left\langle T_c \,\text{exp}\left[ -i \int_c  dt_1 dt_2 \hat \Phi_{\eta j}^\dagger(t_1)\left( e^{i \pi_z \tau_z  \eta(t_1)/2} \hat\Sigma^s(t_1,t_2) e^{-i\pi_z\tau_z  \eta(t_2)/2}\right) \hat \Phi_{\eta j}(t_2)\right] \right\rangle_{H}.
 \end{align}
 Next, we calculate the derivative of the partition function with respect to the counting field $\eta$. We also make  use of $\langle S(\infty)\rangle_{H_l}= T_c \exp \left[ -i\int_{-\infty}^{\infty}dt_1 dt_2 \Phi_{\eta j}^\dagger(t_1) \Sigma^s(t_1-t_2) \Phi_{\eta j}(t_2)\right]$ and keep in mind that  ``$Tr$" is the trace in the Keldysh-Nambu space. Finally, taking the $\eta=0$ limit, we arrive at  the following expression:
 \begin{align}
\langle I_s(t)\rangle =\frac{1}{2} Tr \left[ \tau_z \pi_z  \int_{-\infty}^\infty dt' \left[ G(t,t')\Sigma^s(t',t)-\Sigma^s(t,t')  G(t',t)\right]\right],
\label{ISt}
 \end{align}
 where $ G(t,t')=-i\left\langle T_c \left[ \Phi_{\eta j}(t) \, \Phi_{\eta j}^\dagger(t') \langle S(\infty)\rangle_{H_l}\right] \right\rangle_H$ is the Green function of the full system including the STM and the double-NW setup. We now rotate the Keldysh space such that
 \begin{align}
\tilde G= L \,  G\, L^{-1}~~~~ {\text{and}} ~~~~\tilde \Sigma= L\, \Sigma \, L^{-1}
\Rightarrow  G= L^{-1} \, \tilde G\, L~~~~\,\,\, {\text{and}} ~~~~\hat \Sigma= L^{-1}\, \tilde \Sigma \, L\,.
\end{align}
with 
\begin{align}
\tilde G= \frac{1}{\sqrt 2} \begin{bmatrix}
0 & G_A\\
G_R & G_K
\end{bmatrix}
\,\,\, \text{and}~\tilde \Sigma^s= \frac{1}{\sqrt 2} \begin{bmatrix}
\Sigma_K^s & \Sigma_R^s\\
\Sigma_A^s & 0
\end{bmatrix}.
\end{align}
As a result, we arrive at
\begin{align}
Tr[\pi_z  G(t,t') \Sigma^s(t',t)]&=Tr\pi_z  L^{-1}\tilde G(t,t') L L^{-1}\, \tilde \Sigma^s(t',t) \, L]= Tr[\pi_z L^{-1}\tilde G(t,t') \tilde\Sigma^s(t',t) \, L\,]\nn
&=G_R(t,t')\Sigma^s_K(t',t)+G_K(t,t')\Sigma^s_A(t',t).
\end{align}
Similarly, we calculate
\begin{align}
Tr[\pi_z \Sigma^s(t,t')  G(t',t) ]=\Sigma^s_R(t,t')G_K(t',t)+\Sigma^s_K(t,t')G_A(t',t).
\end{align}
Therefore, the spin  current from Eq. (\ref{ISt}) takes the form
\begin{align}
\langle I_s(t)\rangle=\frac{1}{2}Tr\left( \tau_z \int _{-\infty}^{\infty} dt'\left[G_R(t,t')\Sigma^s_K(t',t)+G_K(t,t')\Sigma^s_A(t',t)-
\Sigma^s_R(t,t')G_K(t',t)-\Sigma^s_K(t,t')G_A(t',t)\right]\right).
\end{align}
Further, the current can be obtained in terms of frequency. To achieve this, we introduce the following double Fourier transformation
\begin{align}
G(t,t')= \sum_{n,m=-\infty}^{\infty} \int_F \frac{d\omega}{2\pi} e^{-i\omega_nt+i\omega_m t'} G_{nm}(\omega)\, ,
\end{align}
where $\omega_n=\omega+nV$ and the integral is performed over a finite domain $F=[0,V]$. For the case of a spin-polarized STM, $G_{nm}(\omega)=\delta_{nm}G(\omega_n)$ and $\Sigma^s_{nm}(\omega)=\delta_{nm}\Sigma^s(\omega_n)$. Also $\sum_n\int_F \frac{d\omega}{2\pi}f(\omega_n)=\int_{-\infty}^{\infty} \frac{d\omega}{2\pi}f(\omega)$. Therefore,
\begin{align}
\langle I_s(\omega_1)\rangle= 2\pi \delta(\omega_1)\frac{1}{2}Tr\Bigg( \tau_z \int_{-\infty}^{\infty} \frac{d\omega}{2\pi} \bigg[G_R(\omega)\Sigma^s_K(\omega)+G_K(\omega)\Sigma^s_A(\omega)-
\Sigma^s_R(\omega)G_K(\omega)-\Sigma^s_K(\omega)G_A(\omega)\bigg]\Bigg).
\end{align} 
The DC current is defined as $\langle I_s(\omega_1)\rangle=2\pi \delta(\omega_1) I^{DC}_s$. Hence,
\begin{align} \label{Ifor}
I^{DC}_s&=\frac{1}{2}Tr\Bigg( \tau_z  \int_{-\infty}^{\infty} \frac{d\omega}{2\pi} \bigg[G_R(\omega)\Sigma^s_K(\omega)+G_K(\omega)\Sigma^s_A(\omega)-
\Sigma^s_R(\omega)G_K(\omega)-\Sigma^s_K(\omega)G_A(\omega)\bigg]\Bigg)\nn
&=\frac{1}{2}Tr\Bigg( \tau_z \int_{-\infty}^{\infty} \frac{d\omega}{2\pi} Re\bigg[G_R(\omega)\Sigma^s_K(\omega)+G_K(\omega)\Sigma^s_A(\omega)\bigg]\Bigg).
\end{align}
In this work, we use  Eq. (\ref{Ifor}) to calculate the spin current  numerically.

\begin{figure*}[t]
\begin{center}\begin{tabular}{cccc}
\epsfig{figure=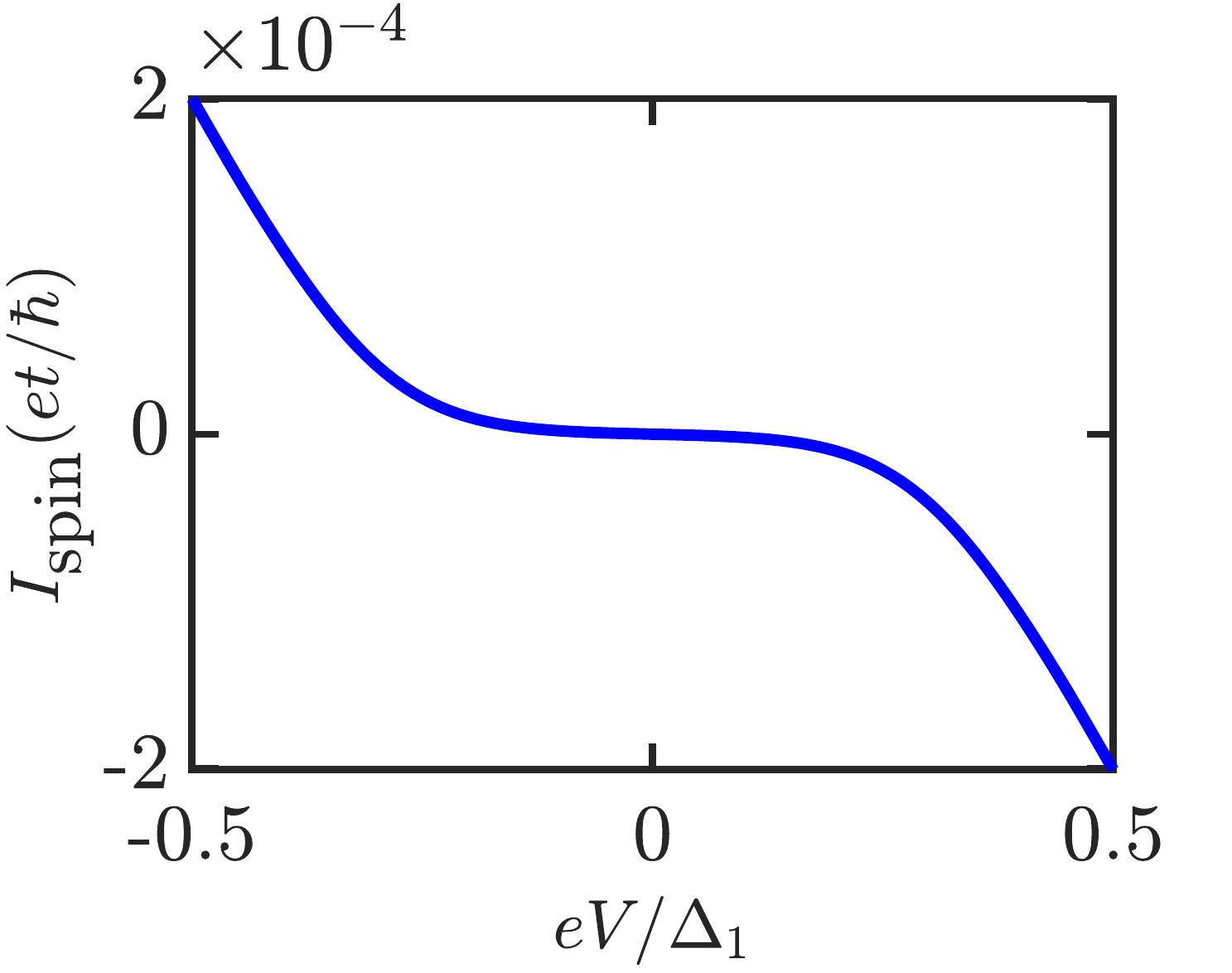,width=1.85in,height=1.38in,clip=true} &\hspace*{-0.4cm}
\raisebox{-0.05cm}{\epsfig{figure=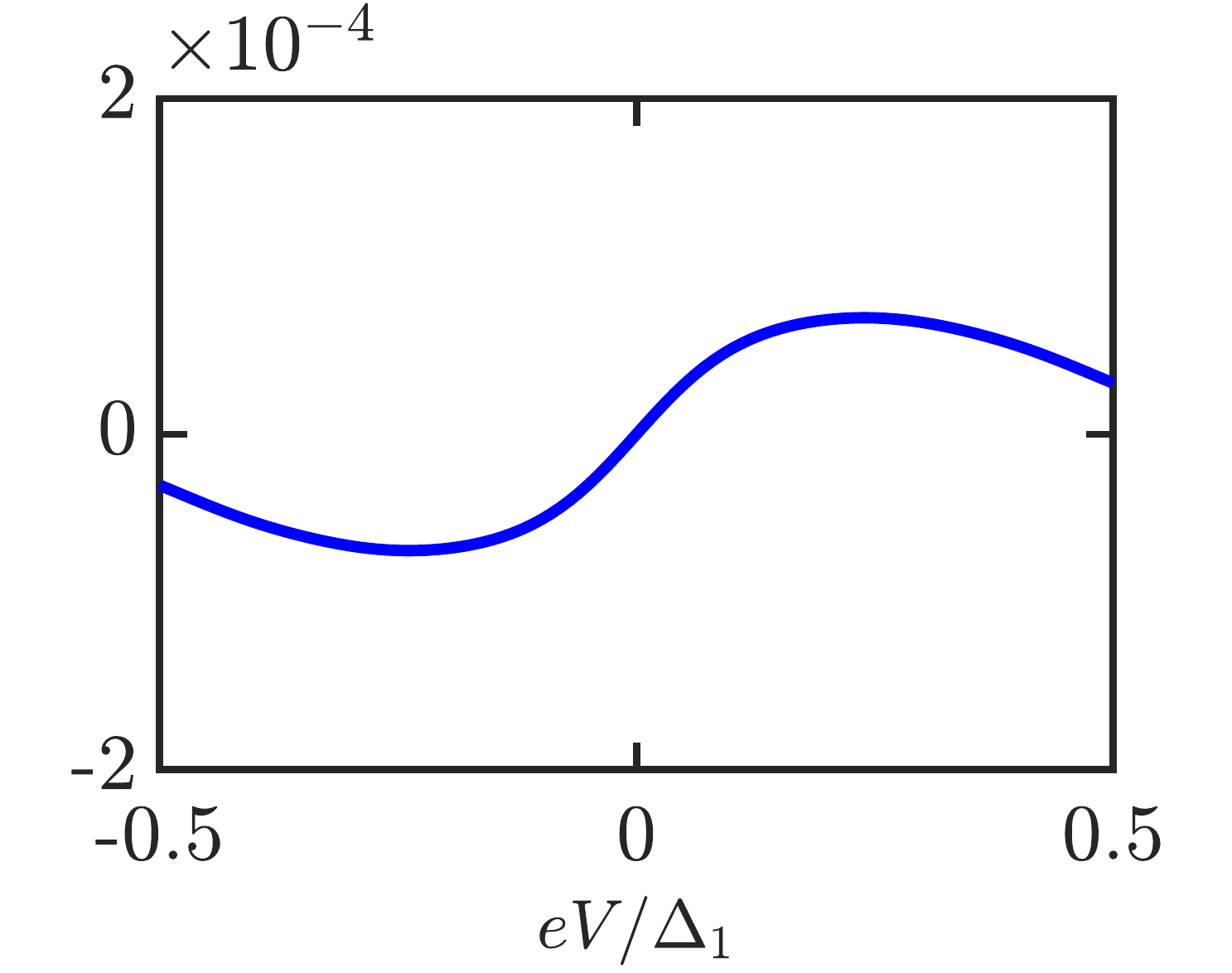,width=1.75in,height=1.4in,clip=true}}
&\hspace*{-0.4cm}
\raisebox{-0.05cm}
{\epsfig{figure=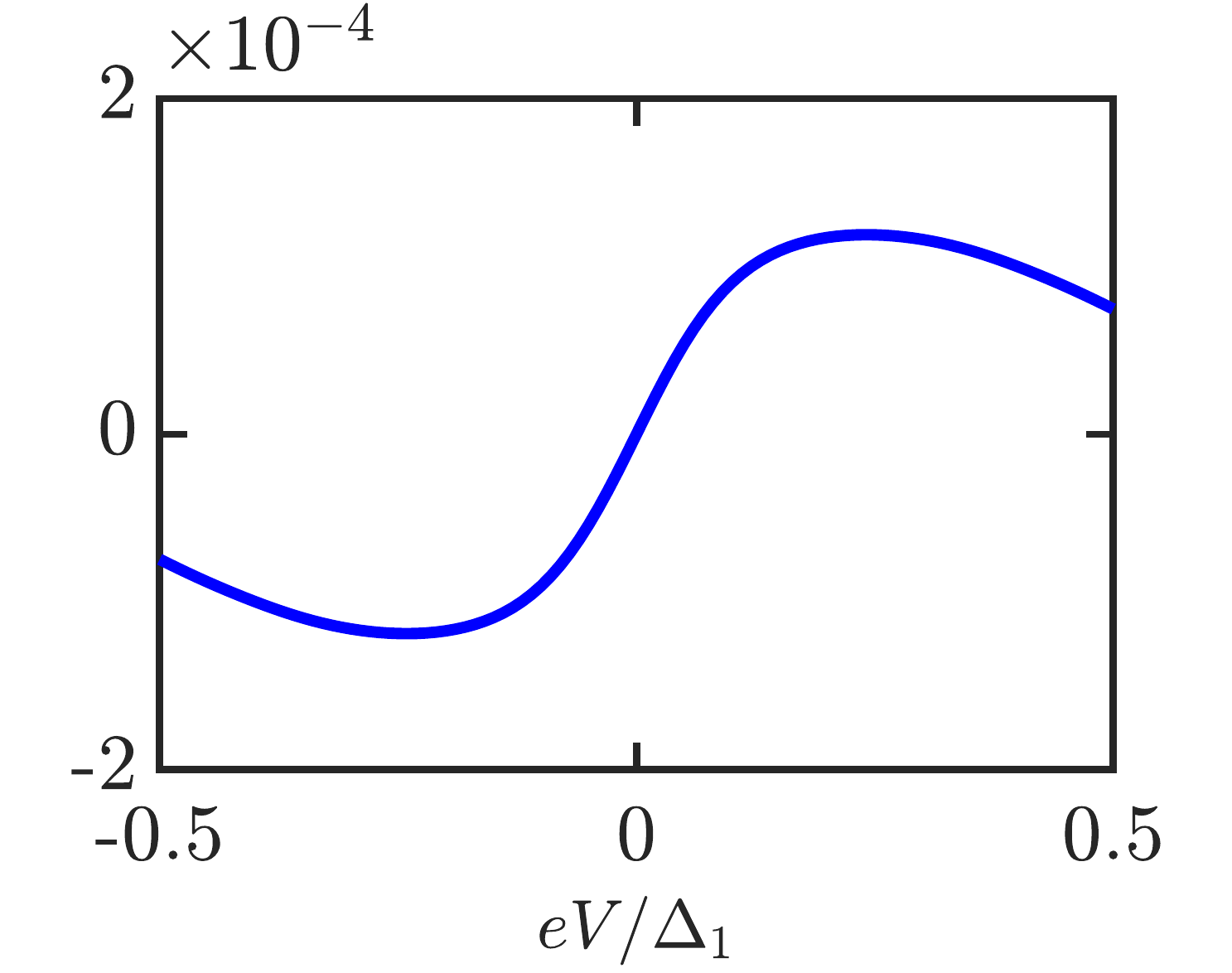,width=1.75in,height=1.39in,clip=true}} &\hspace*{-0.4cm}
\raisebox{-0.09cm}
{\epsfig{figure=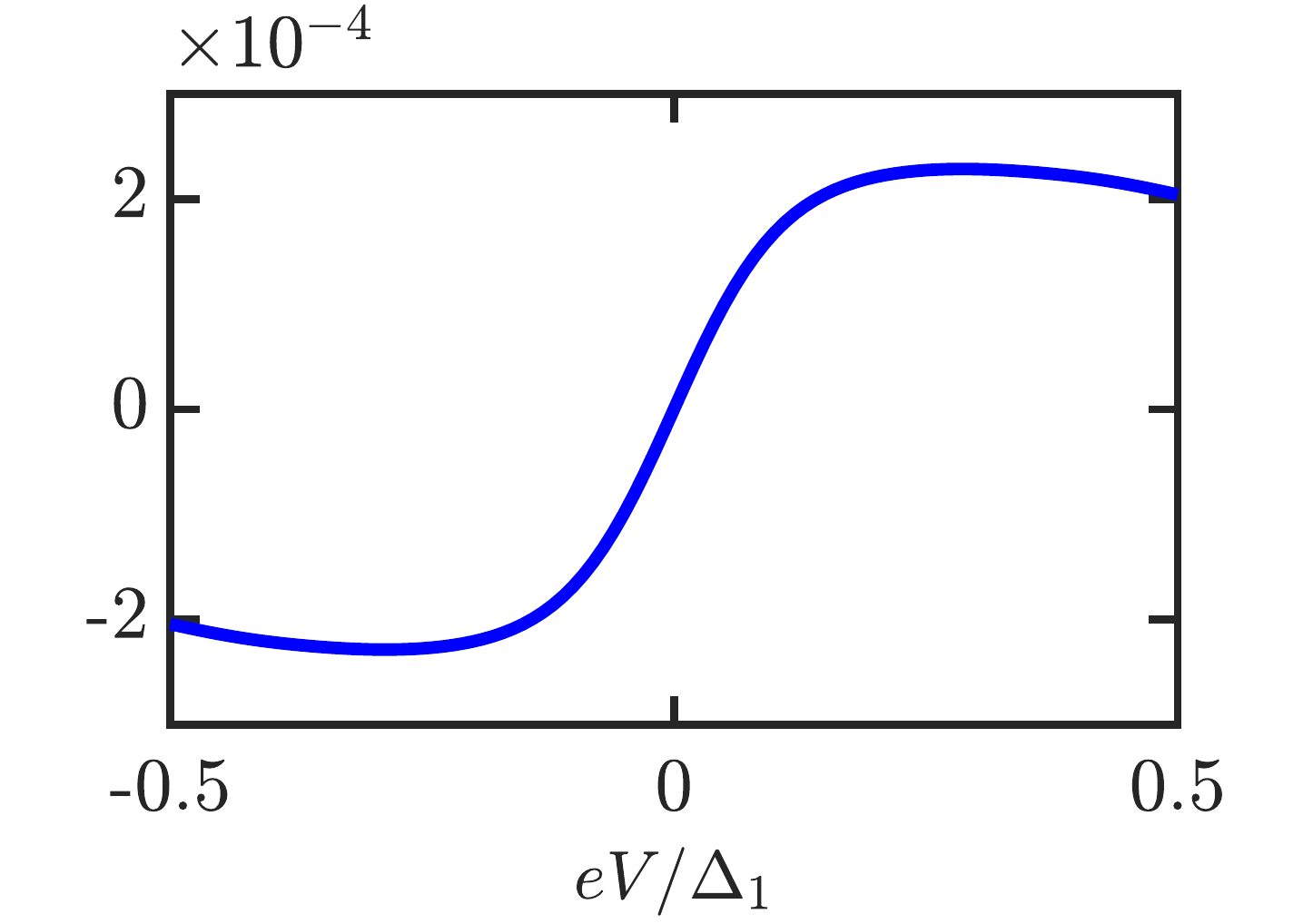,width=1.75in,height=1.42in,clip=true}}\\
\epsfig{figure=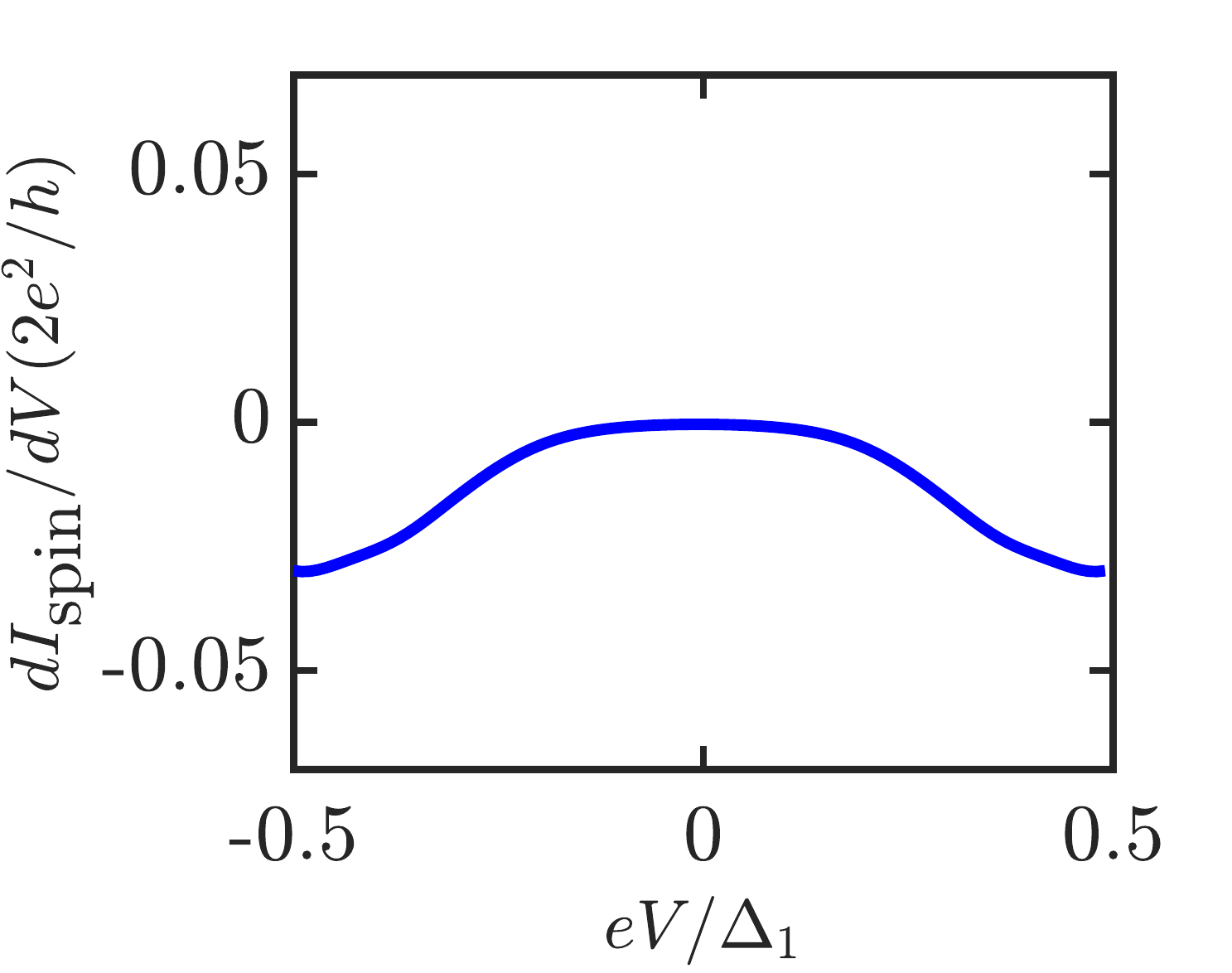,width=1.85in,height=1.38in,clip=true} &\hspace*{-0.4cm}
\raisebox{-0.01cm}{\epsfig{figure=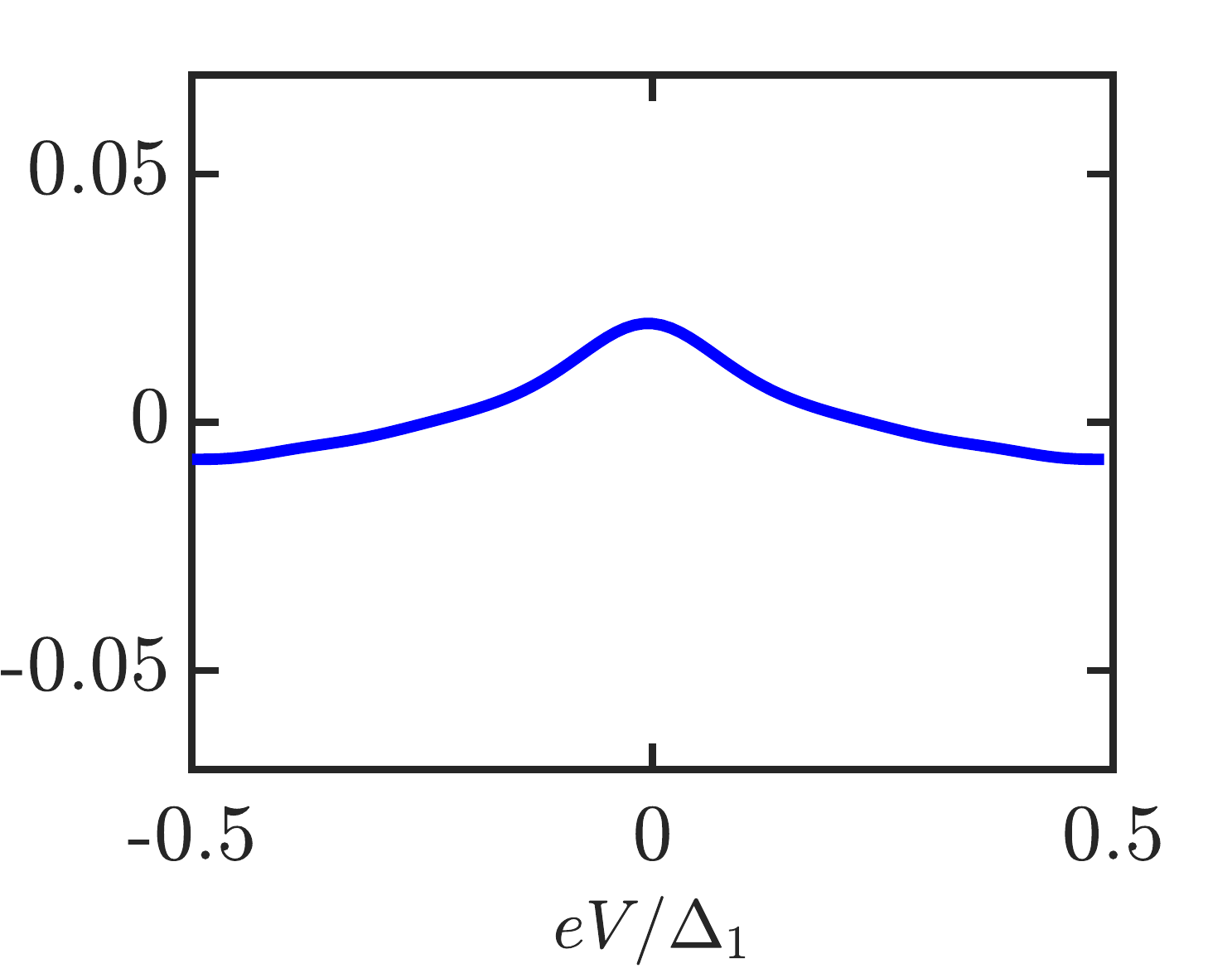,width=1.75in,height=1.38in,clip=true}}
&\hspace*{-0.4cm}
\raisebox{-0.065cm}
{\epsfig{figure=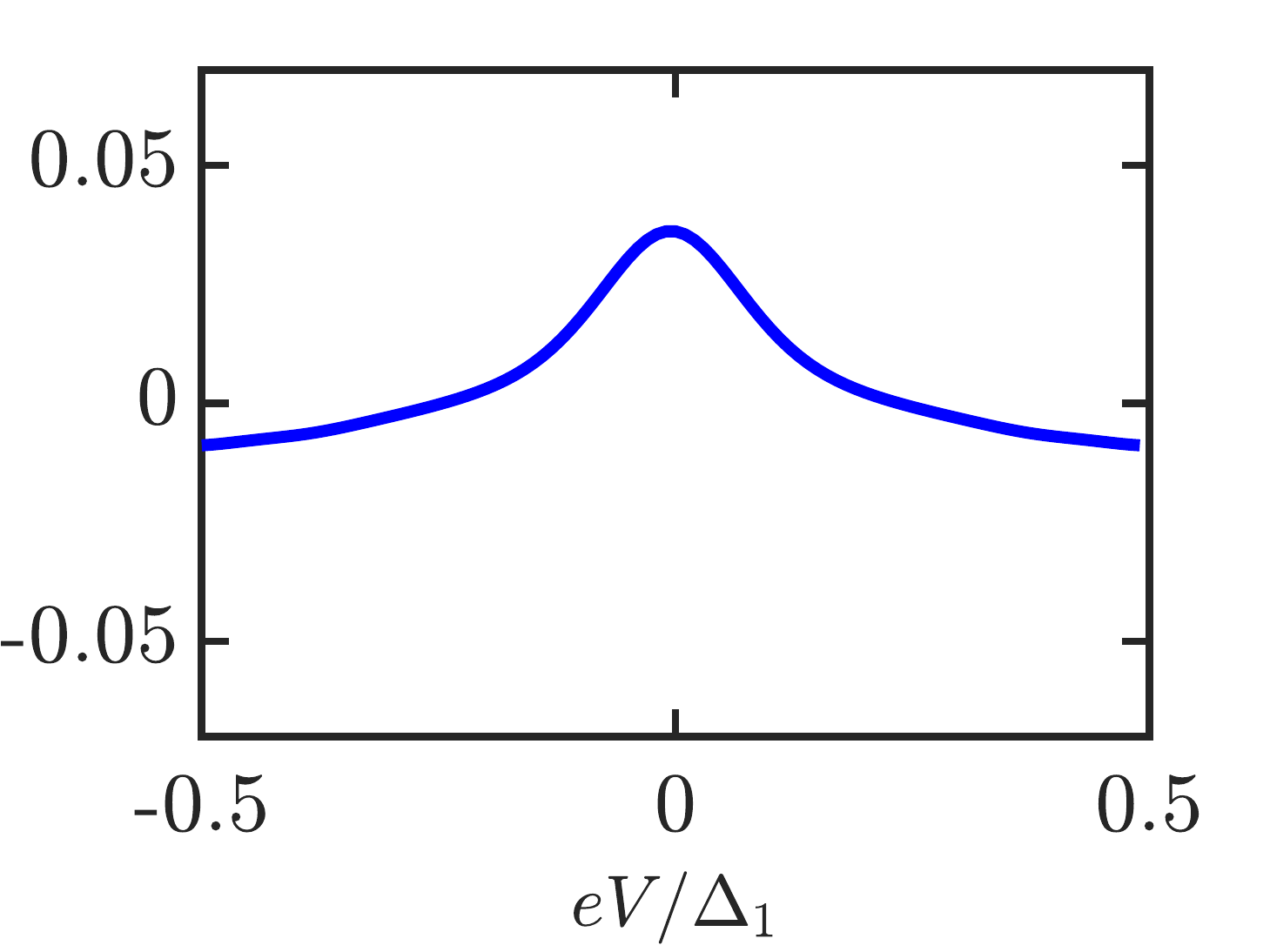,width=1.75in,height=1.4in,clip=true}} &\hspace*{-0.4cm}
\raisebox{-0.08cm}
{\epsfig{figure=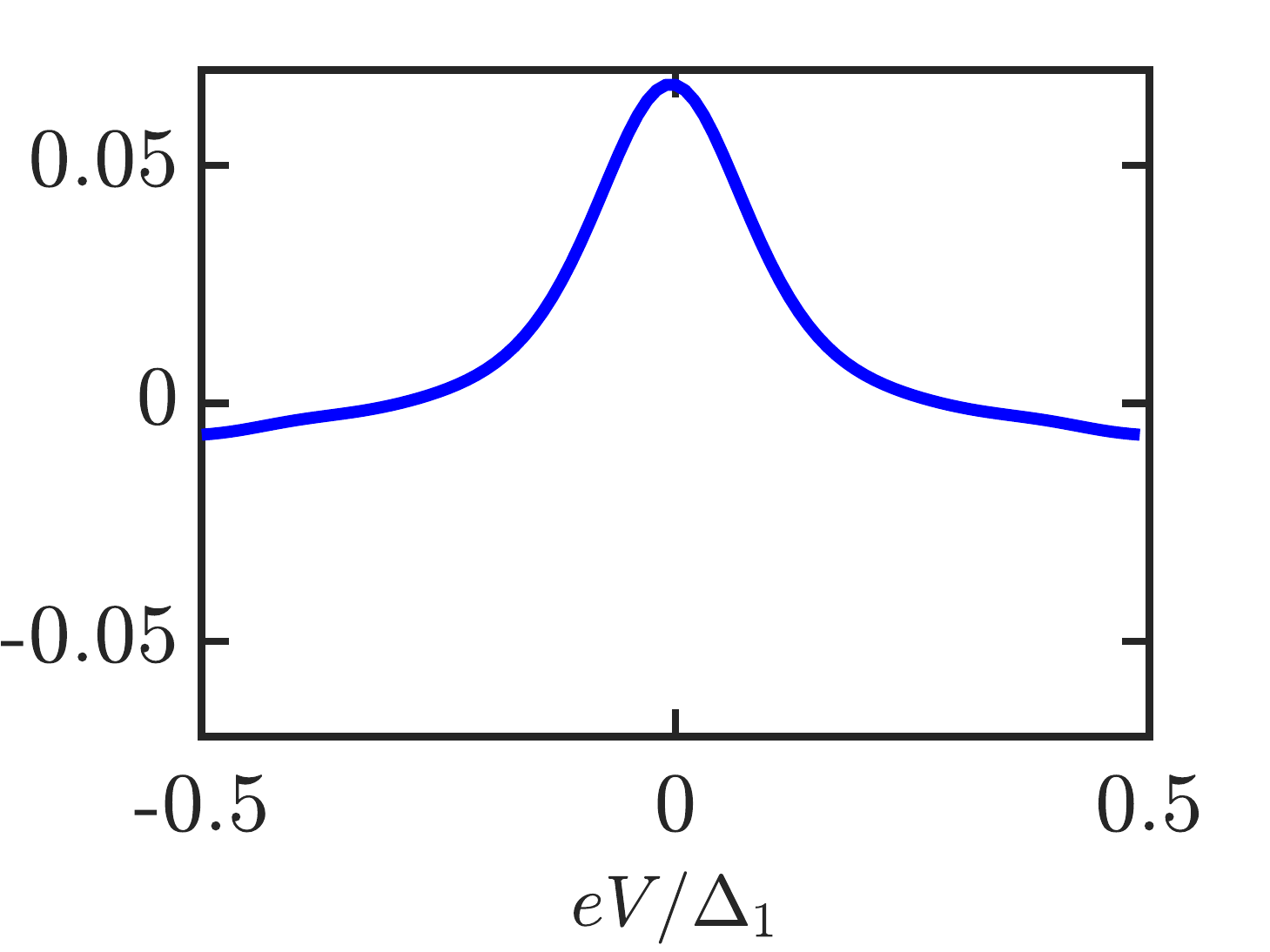,width=1.75in,height=1.41in,clip=true}}
\end{tabular}
\end{center}
\caption{ The same as in Fig. \ref{fig06} but the STM tip is connected at the end of the NW, {\it i.e.} at site $j=5$.   
The spin current and differential conductance show the nonzero contribution of the MBS around zero bias. The sign flip feature is missing as the MBS signal completely masks the bulk contribution responsible for it.}
\label{fig07}
\end{figure*}

\section{Spin Current from MBSs}
In this Appendix, we  consider the case where the SP STM tip is connected to the end of the NW such that the MBS also contributes to the current in the topological phases.  We note that we cannot rely anymore on the spin flip predictions obtained in the momentum space assuming  translation invariance. In such a configuration,  the boundary effects begin to play an important role. The boundary spin that builds up at the NW ends as well as the MBSs prevent us from accessing the bulk properties of the band.
Therefore, the sign flip feature is not captured [see Fig. \ref{fig07}].  

\twocolumngrid

\end{document}